\documentclass[%
 reprint,
 amsmath,amssymb,
 aps,
]{revtex4-1}

\usepackage{graphicx}
\usepackage{dcolumn}
\usepackage{bm}
\usepackage{color}

\newcommand{\aap}{A.\&Ap.}
\newcommand{\mnras}{MNRAS}

\newcommand{\apjl}{ApJ}

\newcommand{\apjs}{ApJS}
\newcommand{\physrep}{Physics Report}
\newcommand{\jcap}{JCAP}

\newcommand{\simgt}{\lower.5ex\hbox{$\; \buildrel > \over \sim \;$}}
\newcommand{\simlt}{\lower.5ex\hbox{$\; \buildrel < \over \sim \;$}}

\begin{document}

\preprint{APS/123-QED}

\title{Perturbation Theory for BAO reconstructed fields: one-loop
  results in real-space matter density field}

\author{Chiaki Hikage}
\affiliation{%
 Kavli Institute for the Physics and Mathematics of the Universe (Kavli IPMU, WPI), University of Tokyo, 5-1-5 Kashiwanoha, Kashiwa, Chiba, 277-8583, Japan
}%
 \email{chiaki.hikage@ipmu.jp}
\author{Kazuya Koyama}%
\affiliation{%
 Institute of Cosmology and Gravitation, University of Portsmouth, Portsmouth PO1 3FX, UK
}%
\author{Alan Heavens}%
\affiliation{%
 Imperial Centre for Inference and Cosmology (ICIC), Department of Physics, Imperial College London, Blackett Laboratory, Prince Consort Road, London SW7 2AZ, UK
}%

\date{\today}

\begin{abstract}   
We compute the power spectrum at one loop order in standard
perturbation theory for the matter density field to which a standard
Lagrangian Baryonic acoustic oscillation (BAO) reconstruction
technique is applied.  The BAO reconstruction method corrects the bulk
motion associated with the gravitational evolution using the inverse
Zel'dovich approximation (ZA) for the smoothed density field. We find
that the overall amplitude of one-loop contributions in the matter
power spectrum substantially decrease after reconstruction. The
reconstructed power spectrum thereby approaches the initial linear
spectrum when the smoothed density field is close enough to linear,
i.e., the smoothing scale $R_s\simgt 10h^{-1}$Mpc. On smaller $R_s$,
however, the deviation from the linear spectrum becomes significant on
large scales ($k\simlt R_s^{-1}$) due to the nonlinearity in the
smoothed density field, and the reconstruction is inaccurate. Compared
with N-body simulations, we show that the reconstructed power spectrum
at one loop order agrees with simulations better than the
unreconstructed power spectrum. We also calculate the tree-level
bispectrum in standard perturbation theory to investigate
non-Gaussianity in the reconstructed matter density field.  We show
that the amplitude of the bispectrum significantly decreases for small
$k$ after reconstruction and that the tree-level bispectrum agrees
well with N-body results in the weakly nonlinear regime.
\end{abstract}

\maketitle

\section{Introduction}
Large-scale structure in our Universe has been widely used for a
variety of cosmological studies. The Baryonic Acoustic Oscillation
(BAO) signature imprinted in the large-scale structure plays a role as
a standard ruler to probe the expansion history of the Universe or
dark energy \citep[e.g.,][]{Eisenstein98}. BAOs have been detected
in various galaxy surveys and used for cosmological studies
\citep[e.g.,][]{Eisenstein05,Cole05,Padmanabhan07,Percival07,Okumura08}.
The broad shape of the power spectrum of evolved density fluctuations
provides a cosmological probe complementary to the Cosmic Microwave Background
\citep{Tegmark04,Cole05,Percival07,Reid10} and a probe of massive
neutrinos \citep[e.g.,][]{Takada06,Saito10}. Both the power
spectrum and the bispectrum are also sensitive to the primordial
non-Gaussianity
\citep[e.g.,][]{Scoccimarro04,Taruya08,Sefusatti09,Desjacques10}.

Nonlinear growth of cosmic large-scale structure and the resulting
bulk flow motion degrade the BAO signals and also biases
the measurements of BAO scales
\cite{Meiksin99,Springel05,Angulo05,SeoEisenstein05,Eisenstein07a,CrocceScoccimarro08}.
The range of scales on which perturbation theory works is limited to large
scales, which makes it difficult to precisely model the evolved matter
density fields \citep[e.g.,][]{Crocce06a,Matsubara08a,Taruya12}. 
Recently there have been renewed interest in extending the range of validity of Perturbation Theory (PT), 
in the context of Effective field theory of Large Scale Structure \cite{eftoflss}. 

Eisenstein et al. \cite{Eisenstein07b} found that the BAO signal was
substantially better recovered by displacing galaxies back to nearly
initial Lagrangian positions. They measured the displacement field
using the inverse Zel'dovich approximation (ZA) applied to the
smoothed density fields.  Ref.~\cite{Padmanabhan09} showed that the
reconstruction reduced the mode-coupling effect and the BAO signature
was much better recovered with a perturbative approach.
Ref.~\citep{Seo10} used N-body simulations to measure the cross
correlation between initial and final density field, known as a
propagator, and found that the reconstructed field recovered the
initial field at higher $k$.  The BAO reconstruction technique has now
been a standard technique of BAO measurements applied to various
galaxy surveys
\citep{Padmanabhan12,Xu13,Anderson14,Tojeiro14,Kazin14,Ross15,Alam16}.
There are a number of planned galaxy surveys such as PFS\citep{PFS},
DESI \citep{DESI16}, HETDEX\citep{HETDEX}, Euclid\citep{Euclid16}, and
WFIRST\citep{WFIRST15} in which the BAO reconstruction technique will
be applied.

What about PT after reconstruction?  Does the range of scales on which
PT works extend to smaller scale than before reconstruction?  To
answer these issues, we derive the exact perturbative formula of the
reconstructed matter power spectrum at one-loop order. Precise
theoretical modeling based on PT is useful for cosmological studies
using reconstructed density field. We use the standard perturbation
theory (SPT) which is a fundamental approach to understand the
nonlinear growth of the density field in weakly nonlinear regime
\citep{Vishniac83,Fry84,Goroff86,SutoSasaki91,Makino92,JainBertschinger94,Bouchet95,ScoccimarroFrieman96,Bernardeau02,JeongKomatsu06}.
We here see how the one-loop order perturbative terms, which are the
leading order of nonlinearity in the matter power spectrum, behave
after reconstruction by varying the smoothing scale $R_s$ for the
smoothed density fields in reconstruction. We also apply our formula
to the formalism of regularized power spectra (RegPT)
\citep{Taruya12,Valageas13}.  Comparing with N-body simulations, we
investigate the valid range of scales on which SPT and RegPT work for
reconstructed matter density field.

We also study the non-Gaussianity in the reconstructed field using the
bispectrum. Since the reconstruction effectively reverses the
growth of the large-scale structure, non-Gaussianity due to the
nonlinear gravity should decrease by reconstruction. We study this
issue by comparing the tree-level PT prediction for the  reconstructed bispectrum with numerical simulations.

Our study improves on previous works in several respects compared with
Padmanabhan et al. and Noh et al. \cite{Padmanabhan09,Noh09}.  We
include several perturbative terms which are necessary to describe the
nonlinearity in the power spectrum.  One is the difference of the
shift field between the data and random. In the standard
reconstruction method developed by \cite{Eisenstein07b}, the positions
of data (e.g., galaxies) are shifted to cancel the effect of bulk flow
and thereby the density field for the data becomes zero at linear
order. To recover the linear-order density field,
uniformly-distributed particles, a so-called ``random'' set, are also
shifted using the same displacement field as used for the data. Since
the data are already displaced from the initial positions, the shift
field for the data should be evaluated at Eulerian positions rather
than Lagrangian positions. On the other hand, the shifts for the
random points should be evaluated at Lagrangian positions. Even though
the same shift field is used for the data and randoms, the difference
of their positions generates additional nonlinearity at the leading
order of PT and changes the bispectrum as well as the power spectrum.
Schmittfull et al. \cite{Schmittfull15} properly took into account the
effect and derived the second-order Eulerian kernel. Another one is
the nonlinearity in the smoothed density field to be used to obtain
the shift field for reconstruction.  In this paper, we explicitly
derive the third-order Eulerian kernel for the reconstructed density
field for the first time and investigate the effect of the
nonlinearity in smoothed density field on the power spectrum. There
has been also no explicit comparison of the perturbative formula of
the reconstructed spectrum with numerical simulations. In this paper,
we study how the one-loop PT works after reconstruction.

This paper is organized as follows: in Section \ref{sec:PT},
perturbative formulae for the power spectrum and bispectrum after
reconstruction are presented. We study how the one-loop terms for the
power spectrum are altered by applying the reconstruction
technique. In Section \ref{sec:results}, we compare our perturbative
formula with N-body simulations and study the range of scales on which
PT works. Section \ref{sec:summary} is devoted to the summary and
conclusions. Throughout the paper, we assume a flat $\Lambda$ CDM
model with the random Gaussian initial condition and use the following
cosmological parameters based on WMAP7+BAO+$H_0$ Mean
\citep{Komatsu11}: $\Omega_b=0.046$, $\Omega_m=0.273$, $n_s=0.963$,
$h=0.704$, $\tau=0.089$, $\sigma_8=0.809$.

\section{Perturbation Theory}
\label{sec:PT}
In this section, we derive the perturbative formula based on the
SPT to describe the nonlinearity in the
real-space matter power spectrum and bispectrum (see also the details
of our derivation in Appendix \ref{sec:app}). We compare the one-loop
order contributions from $\langle\delta^{(1)}\delta^{(3)}\rangle$ and
$\langle\delta^{(2)}\delta^{(2)}\rangle$ before and after
reconstruction. We also apply the 2nd-order Lagrangian Perturbation
Theory (2LPT) instead of Zel'dovich approximation (ZA) in
reconstruction to study the effect on the one-loop terms of the power
spectrum.

\subsection{Standard Perturbation Theory Before Reconstruction}
The Eulerian position of a mass element $\mathbf{x}$ and the
Lagrangian position $\mathbf{q}$ are related through a displacement
field $\mathbf{\Psi(q)}$ as
\begin{equation}
\mathbf{x}=\mathbf{q}+\mathbf{\Psi}(\mathbf{q}). 
\end{equation}
The mass overdensity and its Fourier transform are then written as
\begin{eqnarray}
\delta(\mathbf{x})&=&\int\mathbf{dq}\delta_{\rm D}(\mathbf{x-q-\Psi(q)})-1, \\ 
\tilde\delta_\mathbf{k}&=&\int\mathbf{dq}e^{-i\mathbf{k\cdot
    q}}(e^{-i\mathbf{k\cdot\Psi(q)}}-1),
\label{eq:deltax_unrecon}
\end{eqnarray}
where $\delta_{\rm D}$ is the 3D Dirac delta function.
The displacement field evolves according to the following equation
\begin{equation}
\frac{d^2\mathbf{\Psi}}{dt^2}+2H\frac{d\mathbf{\Psi}}{dt}=-\mathbf{\nabla_x}\phi(\mathbf{q+\Psi(q)}),
\end{equation}
where $H=\dot{a}/a$ is the time-dependent Hubble parameter, and
$\nabla_{\bf x}$ is the spatial derivative with respect to the Eulerian
coordinate. The gravitational potential $\phi$ is determined by the
Poisson equation
\begin{equation}
\mathbf{\nabla_x}^2\phi(\mathbf{x})=4\pi G\bar{\rho}a^2\delta(\mathbf{x}),
\end{equation}
where $G$ is the gravitational constant, $\bar{\rho}$ is the mean
matter density, and $a$ is the scale factor. The velocity field is
assumed to be irrotational throughout this paper. 

In LPT, the displacement field is expanded perturbatively as
\begin{equation}
\mathbf{\Psi}=\mathbf{\Psi}^{(1)}+\mathbf{\Psi}^{(2)}+\mathbf{\Psi}^{(3)}\cdot\cdot\cdot,
\end{equation}
where $\Psi^{(n)}$ has the order of $(\Psi^{(1)})^n$
and each term in Fourier space is given by
\begin{eqnarray}
\label{eq:Lkernel}
\tilde\Psi_\mathbf{k}^{(n)}
=\frac{iD^n(z)}{n!}\int
\frac{\mathbf{dk}_1\cdot\cdot\cdot \mathbf{dk}_n}{(2\pi)^{3n-3}}
\delta_{\rm D} \left(\sum_{j=1}^n\mathbf{k}_j-\mathbf{k}\right) \nonumber \\
\times\mathbf{L}^{(n)}(\mathbf{k}_1,...,\mathbf{k}_n)
\tilde\delta^{\rm L}_{\mathbf{k}_1}
\cdot\cdot\cdot\tilde\delta^{\rm L}_{\mathbf{k}_n},~~~
\end{eqnarray}
where $D(z)$ is the linear growth rate at redshift $z$ normalized by
$D(z=0)=1$ and $\delta^{\rm L}$ is the linear density field at $z=0$.
In the Einstein-de Sitter universe, the Lagrangian perturbative kernels are
time-independent and analytically given by \citep{Catelan95} as
\begin{eqnarray}
\mathbf{L}^{(1)}(\mathbf{k})&=&\frac{\mathbf{k}}{k^2}, \\
\mathbf{L}^{(2)}(\mathbf{k}_1,\mathbf{k}_2)&=&\frac37\frac{\mathbf{k}}{k^2}
(1-\mu_{1,2}^2), 
\end{eqnarray}
and
\begin{eqnarray}
\mathbf{L}^{(3)}(\mathbf{k}_1,&\mathbf{k}_2&,\mathbf{k}_3)
=\frac13\frac{\mathbf{k}}{k^2}
\left[
\frac{5}{7}(1-\mu_{1,2}^2)(1-\mu_{12,3}^2)\right. \nonumber \\
&&-\left.\frac13(1-3\mu_{1,2}^2+2\mu_{1,2}\mu_{2,3}\mu_{3,1})+{\rm (2~perm.)}\right]
\nonumber \\
&&+\mathbf{k\times T}(\mathbf{k}_1,\mathbf{k}_2,\mathbf{k}_3),
\end{eqnarray}
where $\mathbf{k=k}_1+\cdot\cdot\cdot+\mathbf{k}_n$,
$\mu_{i,j}=\mathbf{k}_i\cdot \mathbf{k}_j/k_ik_j$, and
$\mu_{ij,k}=(\mathbf{k}_i+\mathbf{k}_j)\cdot \mathbf{k}_k/
|\mathbf{k}_i+\mathbf{k}_j|k_k$. The transverse part $\mathbf{k\times
  T}$ is unnecessary in the following application \citep{Catelan95}.
In more general cosmologies, the above expression is still a good
approximation \citep{Bernardeau02}, while the exact formula is given
by \cite{Fry84,Goroff86,JainBertschinger94}. 

The perturbation series of the matter overdensity $\delta$ is given as
\begin{equation}
\tilde\delta^{(1)}_\mathbf{k}=-i\mathbf{k\cdot\tilde\Psi}^{(1)}_\mathbf{k}
=D(z)\tilde\delta^{\rm L}_\mathbf{k}, 
\end{equation}
\begin{eqnarray}
\tilde\delta_\mathbf{k}^{(n)} &=& D^n(z)\int
\frac{\mathbf{dk}_1\cdot\cdot\cdot \mathbf{dk}_n}{(2\pi)^{3n-3}}
\delta_{\rm D}\left(\sum_{j=1}^n\mathbf{k}_j-\mathbf{k}\right) \nonumber \\
&& \times F_n(\mathbf{k}_1,\cdot\cdot\cdot,\mathbf{k}_n)
\tilde\delta^{\rm L}_{\mathbf{k}_1}\cdot\cdot\cdot
\tilde\delta^{\rm L}_{\mathbf{k}_n},
\end{eqnarray}
where $F_n$ is the $n$-th order Eulerian perturbative kernel.
Following the previous literature
\citep[e.g.,][]{Goroff86,JainBertschinger94}, the symmetrization
factor $1/n!$ is not included in the definition of the Eulerian kernel,
while it is included in that of the Lagrangian kernel
(eq.[\ref{eq:Lkernel}]).  The second- and third-order Eulerian kernels
are given by \cite{Fry84,Goroff86,JainBertschinger94}
\begin{eqnarray}
\label{eq:F2_unrecon}
F_2(\mathbf{k}_1,\mathbf{k}_2)&=&\frac12
[\mathbf{k}\cdot \mathbf{L}^{(2)}(\mathbf{k}_1,\mathbf{k}_2)
+(\mathbf{k}\cdot \mathbf{L}^{(1)}(\mathbf{k}_1))
(\mathbf{k}\cdot \mathbf{L}^{(1)}(\mathbf{k}_2))], \nonumber \\
&=&\frac57+\frac27\left(\frac{\mathbf{k}_1\cdot \mathbf{k}_2}{k_1k_2}\right)^2
+\frac{\mathbf{k}_1\cdot \mathbf{k}_2}{2k_1k_2}\left(\frac{k_2}{k_1}+\frac{k_1}{k_2}\right),
\end{eqnarray}
and
\begin{eqnarray}
F_3(\mathbf{k}_1,&\mathbf{k}_2&,\mathbf{k}_3)=
\frac{1}{6}\left[\mathbf{k}\cdot \mathbf{L}^{(3)}
(\mathbf{k}_1,\mathbf{k}_2,\mathbf{k}_3)\right.  \nonumber \\
&&+(\mathbf{k}\cdot \mathbf{L}^{(1)}(\mathbf{k}_1))
(\mathbf{k}\cdot \mathbf{L}^{(1)}(\mathbf{k}_2))
(\mathbf{k}\cdot \mathbf{L}^{(1)}(\mathbf{k}_3))
\nonumber \\
&& + \left.\left\{(\mathbf{k}\cdot \mathbf{L}^{(1)}(\mathbf{k}_1))
(\mathbf{k}\cdot \mathbf{L}^{(2)}(\mathbf{k}_2,\mathbf{k}_3))+{\rm (2~perms.)}
\right\}\right]. \nonumber \\
\end{eqnarray}

The matter power spectrum $P(k)$ is defined as
\begin{equation}
\langle\tilde\delta_\mathbf{k}\tilde\delta_\mathbf{k'}\rangle
\equiv (2\pi)^3\delta_{\rm D}(\mathbf{k+k'})P(k).
\end{equation}
The power spectrum up to one-loop order is given by
\begin{equation}
\label{eq:stdpk}
P(k,z)=D^2(z)P_{11}(k)+D^4(z)[P_{22}(k)+P_{13}(k)],
\end{equation}
where $P_{nm}$ is the contribution from
$\langle\tilde\delta_\mathbf{k}^{(n)}\tilde\delta_\mathbf{k}^{(m)}\rangle$.
The leading order term is the linear power spectrum
\begin{equation}
P_{11}(k)=P_{\rm L}(k),
\end{equation}
and the next-leading order terms are the following two one-loop terms: 
\begin{eqnarray}
P_{22}(k)&=&2\int\frac{\mathbf{dp}}{(2\pi)^3}P_{\rm L}(|\mathbf{k-p}|)P_{\rm L}(p)[F_2(\mathbf{k-p},\mathbf{p})]^2, \nonumber \\
&=&\frac{1}{98}\frac{k^3}{4\pi^2}\int_0^\infty dr P_L(kr)
\int_{-1}^1 dx P_L(k_\ast) \nonumber \\
&& \times \frac{(3r+7x-10rx^2)^2}{(1+r^2-2rx)^2},
\end{eqnarray}
where $k_\ast\equiv k(1+r^2-2rx)^{1/2}$ and
\begin{eqnarray}
P_{13}(k)&=&6P_{\rm L}(k)\int\frac{\mathbf{dp}}{(2\pi)^3}P_{\rm L}(p)
F_3(\mathbf{k},\mathbf{p},\mathbf{-p}), \nonumber \\
&=&\frac{1}{252}\frac{k^3}{4\pi^2}P_L(k)\int_0^{\infty}drP_L(kr)
\left[\frac{12}{r^2}-158+100r^2\right. \nonumber \\
&&-\left.42r^4+\frac{3}{r^3}(r^2-1)^3(7r^2+2)\ln\left|\frac{1+r}{1-r}\right|\right].
\label{eq:P13_unrecon}
\end{eqnarray}
The bispectrum at the tree-level order is given by
\begin{equation}
\label{eq:bis_unrecon}
B(k_1,k_2,k_3)=2F_2(\mathbf{k}_1,\mathbf{k}_2)D^4(z)
P_{\rm L}(k_1)P_{\rm L}(k_2)+({\rm 2~perms.}).
\end{equation}

\subsection{Standard Perturbation Theory After Reconstruction}
Next we move to the reconstructed field. To reconstruct the density
field, we shift the observed particle positions to correct the bulk
flow motion following the procedure of \cite{Eisenstein07b}. When the
shift field $\mathbf{s}$ is computed from the negative ZA
\citep{Zeldovich70} of the smoothed density field, it is obtained by
\begin{eqnarray}
\label{eq:sk_ZA}
\mathbf{\tilde{s}_k}&=&-iW(k)\mathbf{L}^{(1)}(\mathbf{k})\tilde\delta_\mathbf{k},
\end{eqnarray}
where $W(k)$ is the smoothing kernel. We adopt a Gaussian kernel
$W(k)=\exp(-k^2R_s^2/2)$, varying the smoothing scale of $R_s$
throughout the paper. Since the shift field is computed from the
negative ZA of the smoothed density field (eq.\ref{eq:sk_ZA}), the
perturbative series of the shift field is given by that of the
smoothed density field as
\begin{eqnarray}
\label{eq:skn_ZA}
\mathbf{\tilde{s}_k}^{(n)}=-iW(k)\mathbf{L}^{(1)}(\mathbf{k})\tilde\delta_\mathbf{k}^{(n)}.
\end{eqnarray}
This can be rewritten in a similar form to the Lagrangian kernel (eq. \ref{eq:Lkernel}) as
\begin{eqnarray}
\mathbf{\tilde{s}_k}^{(n)}&=&\frac{iD^n(z)}{n!}
\int\frac{\mathbf{dk}_1\cdot\cdot\cdot \mathbf{dk}_n}{(2\pi)^{3n-3}} 
\delta_{\rm D}\left(\sum_{j=1}^n \mathbf{k}_j-\mathbf{k}\right) 
\nonumber \\
&&\times
\mathbf{S}^{(n)}(\mathbf{k}_1,...,\mathbf{k}_n)
\tilde\delta^{\rm L}_{\mathbf{k}_1}\cdot\cdot\cdot\tilde\delta^{\rm L}_{\mathbf{k}_n},
\end{eqnarray}
where the kernel of the shift field $\mathbf{S}^{(n)}$ is given by
\begin{equation}
\label{eq:shiftkernel}
\mathbf{S}^{(n)}(\mathbf{k}_1,...,\mathbf{k}_n)=
-n!W(k)\mathbf{L}^{(1)}(\mathbf{k})F_n(\mathbf{k}_1,...,\mathbf{k}_n).
\end{equation}
At linear order, $\mathbf{S}^{(1)}$ becomes $-W(k)\mathbf{L}^{(1)}$
and thereby the displacement due to gravitational evolution is
canceled out by reconstruction on large scales ($W(k)\simeq 1$), i.e.,
$\mathbf{\tilde{\Psi}_k}^{(1)}+\mathbf{\tilde{s}_k}^{(1)}\simeq 0$.
At higher order, however, they are not completely canceled out with
each other even on large scales.

The displaced density field is written as
\begin{eqnarray}
\label{eq:deltak_unrecon}
\tilde\delta_\mathbf{k}^{\rm (d)}=\int\mathbf{dq}e^{-i\mathbf{k\cdot q}}(e^{-i\mathbf{k\cdot[\Psi(q)+s(x)]}}-1),
\end{eqnarray}
where the shift field of the evolved mass particles is evaluated at
the Eulerian positions $\mathbf{s(x)}$ rather than at the Lagrangian
position $\mathbf{s(q)}$, as discussed in \cite{Schmittfull15}.  The
difference of the shift field between the Eulerian positions
$\mathbf{x}$ and Lagrangian positions $\mathbf{q}$ is perturbatively
expanded in terms of $\mathbf\Psi$ as
\begin{eqnarray}
\mathbf{s(x)}&=&\int \frac{\mathbf{dk}}{(2\pi)^3}~\mathbf{\tilde{s}_k}~e^{i\mathbf{k\cdot (q+\Psi(q))}}, \\
&=&\sum_{n=0}^\infty\int \frac{\mathbf{dk}}{(2\pi)^3}~\mathbf{\tilde{s}_k}~e^{i\mathbf{k\cdot q}} \left[\frac{1}{n!}(i\mathbf{k\cdot \Psi(q)})^n\right], \nonumber \\
\label{eq:sx_pb}
&=&\mathbf{s(q)}+\mathbf{(\Psi(q)\cdot\nabla)s(q)}+\frac{1}{2}\mathbf{(\Psi(q)\cdot\nabla)}^2\mathbf{s(q)}\cdot\cdot\cdot . \nonumber \\
\end{eqnarray}
The shifted density field of a spatially uniform grid or random
is given by
\begin{equation}
\tilde\delta_\mathbf{k}^{\rm (s)}
=\int\mathbf{dq}e^{-i\mathbf{k\cdot q}}(e^{-i\mathbf{k\cdot s(q)}}-1),
\end{equation}
where the shift field of the (unevolved) uniform grid is evaluated at
the Lagrangian position. The reconstructed density field is given as
\begin{eqnarray}
\label{eq:reconfield}
\tilde\delta_\mathbf{k}^{\rm (rec)}&\equiv &
\tilde\delta_\mathbf{k}^{\rm (d)} - \tilde\delta_\mathbf{k}^{\rm (s)} \nonumber \\
&=&\int \mathbf{dq}e^{-i\mathbf{k\cdot q}}
e^{-i\mathbf{k\cdot s(q)}}(e^{-i\mathbf{k\cdot[\Psi(q)+s(x)-s(q)]}}-1). \nonumber \\
\end{eqnarray}

At linear order, the reconstructed density field is unaltered as
\begin{eqnarray}
\label{eq:delrec1}
\delta_\mathbf{k}^{\rm (rec) (1)}&=&\delta_\mathbf{k}^{(1)}.
\end{eqnarray}
There is a difference in higher-order terms of $\delta^{\rm (rec)}$ 
\begin{eqnarray}
\label{eq:F_n}
\tilde\delta_\mathbf{k}^{\rm (rec) (n)}&=&D^n(z)
\int\frac{\mathbf{dk}_1\cdot\cdot\cdot \mathbf{dk}_n}{(2\pi)^{3n-3}}\delta_{\rm D}
\left(\sum_{j=1}^n \mathbf{k}_j-\mathbf{k}\right) \nonumber \\
&&\times F_n^{\rm (rec)}(\mathbf{k}_1,...,\mathbf{k}_n)
\tilde\delta^{\rm L}_{\mathbf{k}_1}\cdot\cdot\cdot\tilde\delta^{\rm L}_{\mathbf{k}_n},
\end{eqnarray}
where $F_n^{\rm (rec)}$ is the Eulerian kernel for the reconstructed
matter density field. In this paper, we derive $F_n^{\rm (rec)}$ up to
third order by expanding the reconstructed density fields
(eq. \ref{eq:reconfield}) perturbatively.  The detail of derivations is summarized in Appendix \ref{sec:app}. The second-order
Eulerian kernel for the reconstructed field $F_2^{\rm (rec)}$ becomes
\begin{eqnarray}
F_2^{\rm (rec)}(&\mathbf{k}_1&,\mathbf{k}_2)=F_2(\mathbf{k}_1,\mathbf{k}_2) \nonumber \\
&&+\frac12\left[(\mathbf{k}\cdot \mathbf{S}^{\rm (1)}(\mathbf{k}_1))
+(\mathbf{k}\cdot \mathbf{S}^{\rm (1)}(\mathbf{k}_2))\right], \nonumber \\
&=&\frac57-\frac{W_1+W_2}{2}+\frac27
\left(\frac{\mathbf{k}_1\cdot \mathbf{k}_2}{k_1k_2}\right)^2 \nonumber \\
&&+\left(\frac{\mathbf{k}_1\cdot \mathbf{k}_2}{2k_1k_2}\right)
\left[\frac{k_2}{k_1}(1-W_1)+\frac{k_1}{k_2}(1-W_2)\right],
\nonumber \\
\label{eq:F2_recon}
\end{eqnarray}
where $F_n$ without upperscript of (rec) denotes the $n$-th order
Eulerian kernel without reconstruction (eq. \ref{eq:F2_unrecon}) and
$W_{i}\equiv W(k_i)$. The third-order kernel becomes
\begin{eqnarray}
F_3^{\rm (rec)}(&\mathbf{k}_1&,\mathbf{k}_2,\mathbf{k}_3)=
F_3(\mathbf{k}_1,\mathbf{k}_2,\mathbf{k}_3) \nonumber \\
&&+\frac16\left[2(\mathbf{k\cdot S}^{(1)}(\mathbf{k}_1))F_2(\mathbf{k}_2,\mathbf{k}_3)\right.
\nonumber \\
&&+(\mathbf{k\cdot S}^{(1)}(\mathbf{k}_1))(\mathbf{k\cdot S}^{(1)}(\mathbf{k}_2))
\nonumber \\
&& \left.+(\mathbf{k\cdot S}^{(2)}(\mathbf{k}_1,\mathbf{k}_2))+{\rm (2~perms.)}
\right],
\label{eq:F3_recon}
\end{eqnarray}
where $\mathbf{k}_{ij}\equiv\mathbf{k}_i+\mathbf{k}_j$ and
$W_{ij}\equiv W(|\mathbf{k}_{ij}|)$. The last term including the
$\mathbf{S}^{(2)}$ kernel comes from the second-order nonlinearity of
the shift field $\mathbf{s(x)}$.

At linear order, the power spectrum is unchanged by reconstruction as
shown in the equation (\ref{eq:delrec1}). The one-loop terms for the
matter power spectrum are altered as
\begin{eqnarray}
\label{eq:P22_rec}
P_{22}^{\rm (rec)}&(k)&=2\int\frac{\mathbf{dp}}{(2\pi)^3}P_{\rm L}(|\mathbf{k-p}|)P_{\rm L}(p)[F_2^{\rm (rec)}(\mathbf{k-p},\mathbf{p})]^2, \nonumber \\
&=&\frac{1}{98}\frac{k^3}{4\pi^2}\int_0^\infty dr P_L(kr)\int_{-1}^1 dx 
P_L(k_\ast) \nonumber \\
&&
\times \left[\frac{3r+7x-10rx^2-7W_{k_\ast}r(1-rx)}{1+r^2-2rx}
-7W_{kr}x \right]^2, \nonumber \\
\end{eqnarray}
where $k_\ast=k(1+r^2-2rx)^{1/2}$ and
\begin{eqnarray}
\label{eq:P13_rec}
P_{13}^{\rm (rec)}&(k)&=6P_{\rm L}(k)\int\frac{\mathbf{dp}}{(2\pi)^3}P_{\rm L}(p)
F_3^{\rm (rec)}(\mathbf{k},\mathbf{p},\mathbf{-p}), \nonumber \\
&=&P_{13}(k)+\frac{k^3}{4\pi^2}P_L(k)\int_0^{\infty}drP_L(kr) \nonumber \\
&\times & \left[\frac{2}{3}(2W_{kr}-W_{kr}^2+2r^2W_{kr})
\right. \nonumber \\
&-&
\int^{1}_{-1}dx
\left.
\frac{2r(1-rx)(10r+4rx^2-7r^2x-7x)}{7(1+r^2-2rx)}W_{k_\ast}\right].
\nonumber \\
\end{eqnarray}

The tree-level bispectrum for the reconstructed matter density field
is also obtained by replacing $F_2$ with $F_2^{\rm (rec)}$ as
\begin{eqnarray}
\label{eq:bis_recon}
B(k_1,k_2,k_3)&=&2F_2^{\rm (rec)}(\mathbf{k}_1,\mathbf{k}_2)D^4(z)
P_{\rm L}(k_1)P_{\rm L}(k_2) \nonumber \\
&&+({\rm 2~perms.}).
\end{eqnarray}

\subsection{Power Spectrum Before and After reconstruction}
\begin{figure*}
\begin{center}
\includegraphics[width=7.5cm]{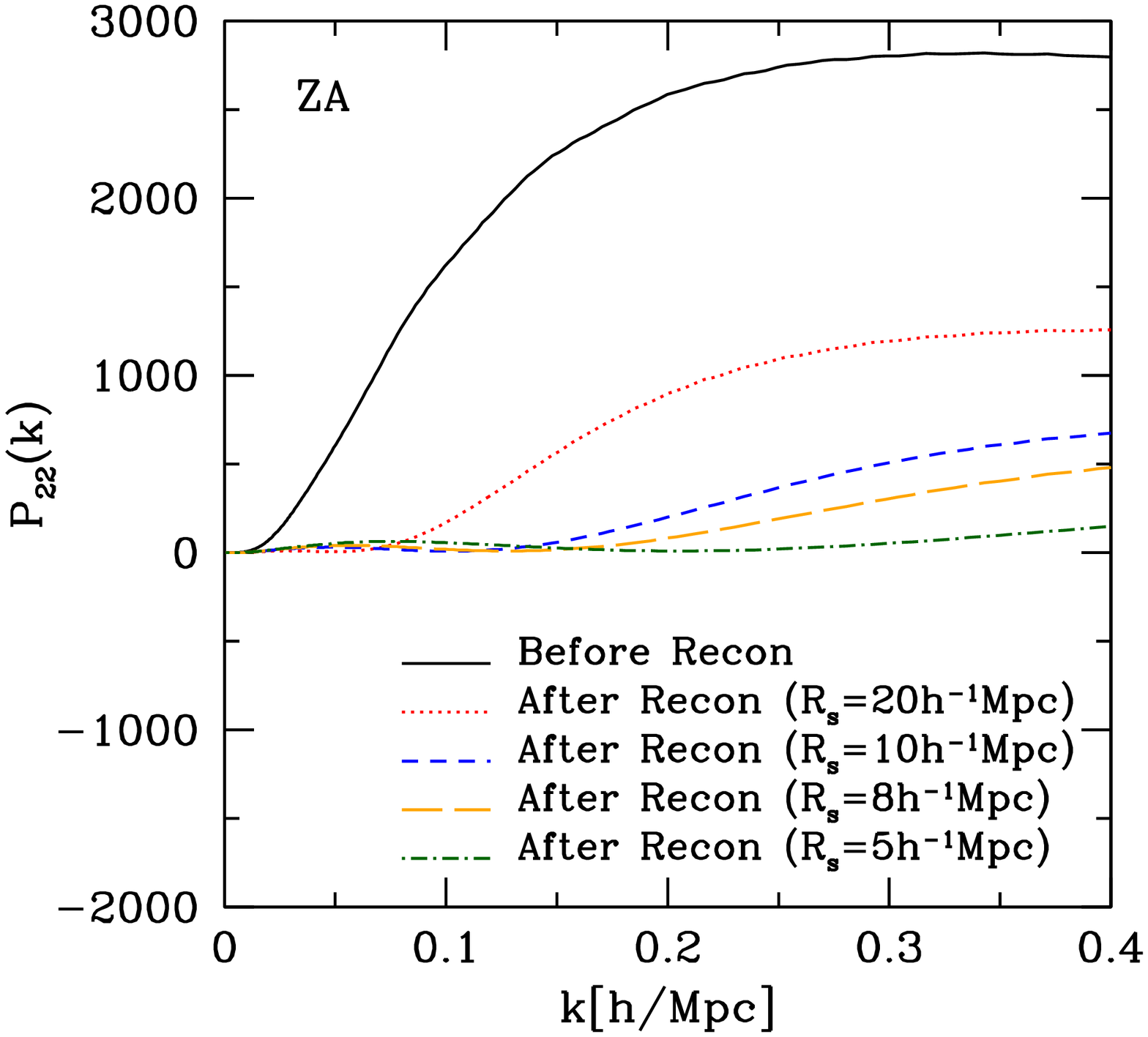}
\includegraphics[width=7.5cm]{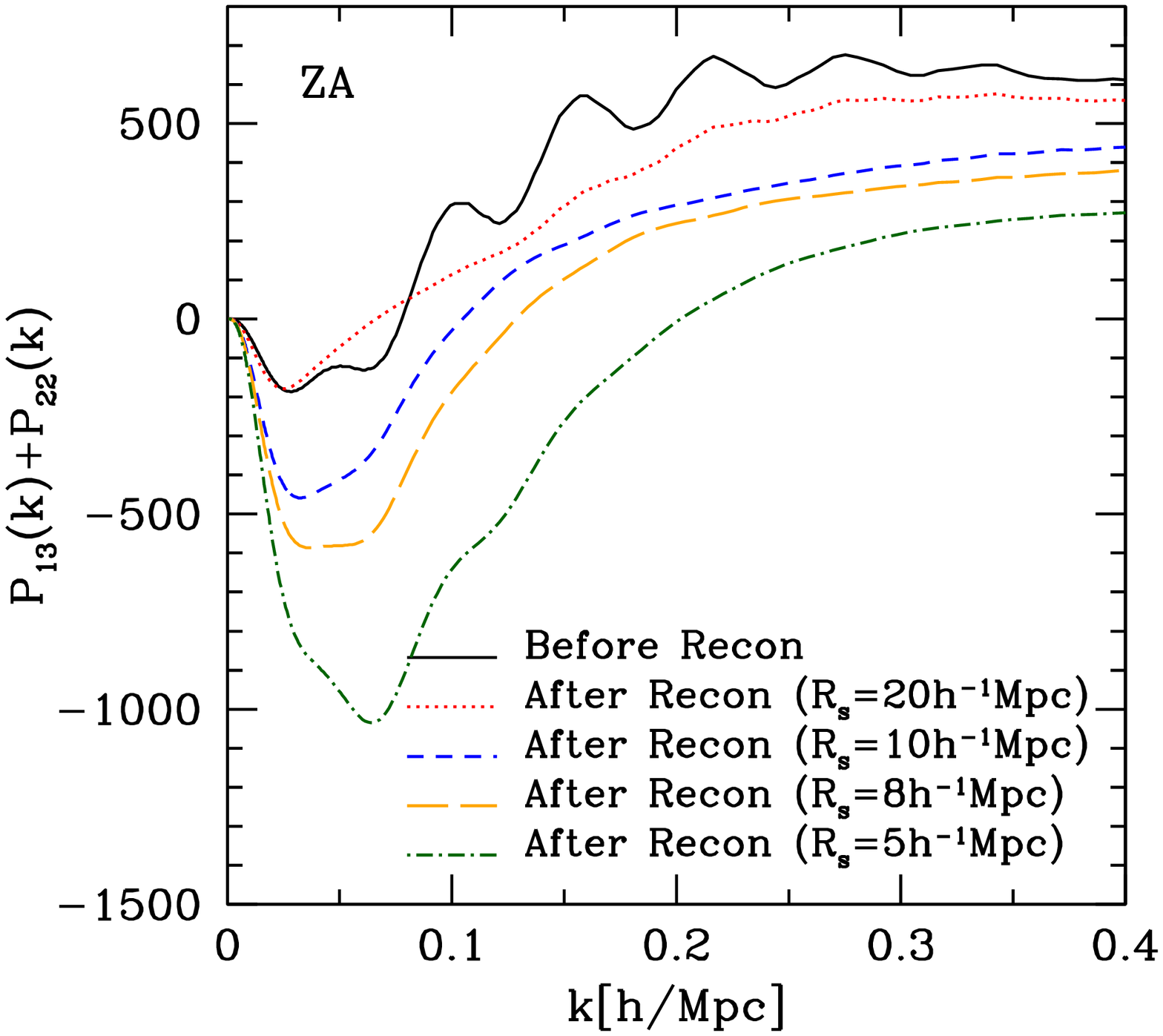}
\includegraphics[width=7.5cm]{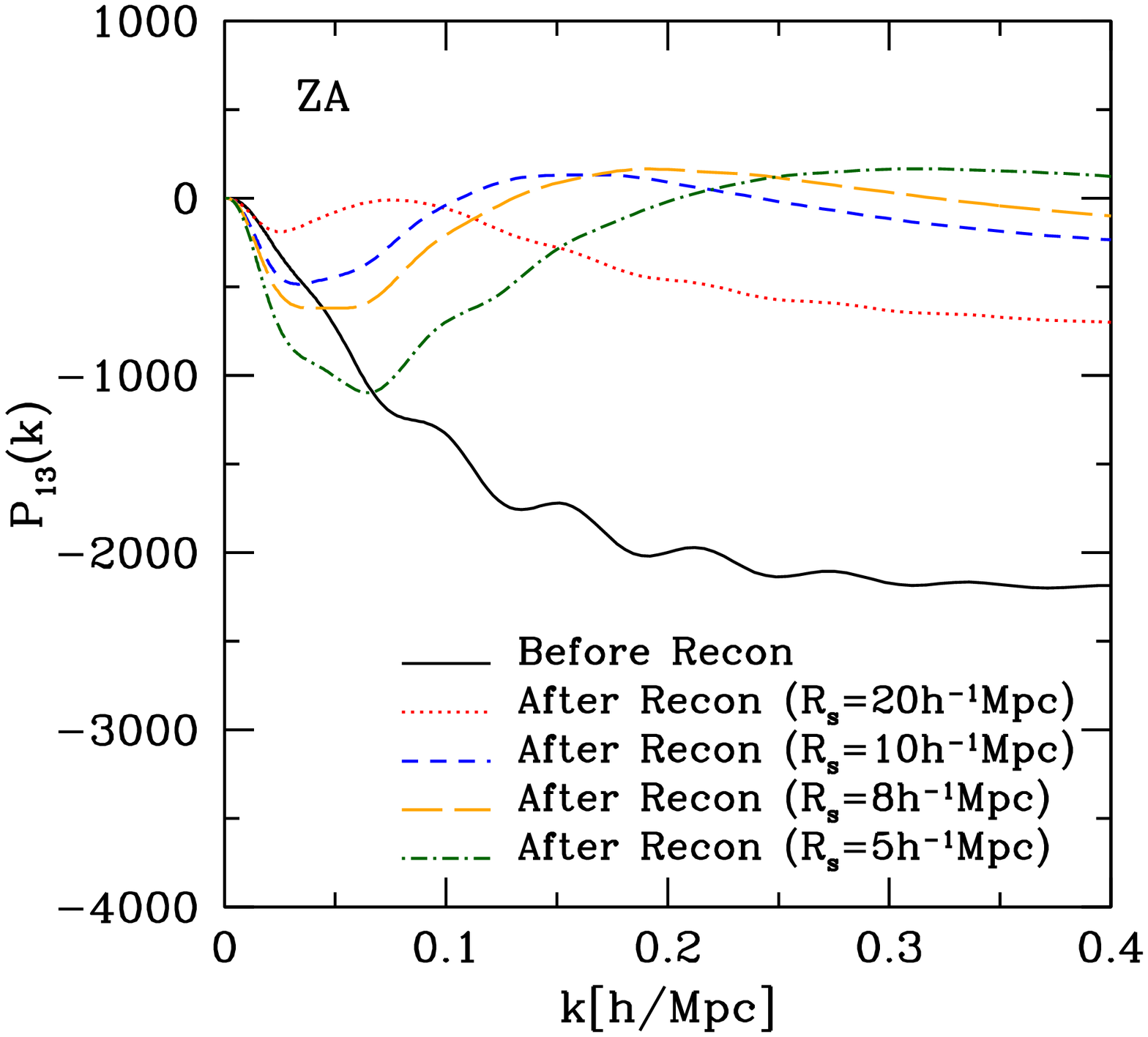}
\includegraphics[width=7.5cm]{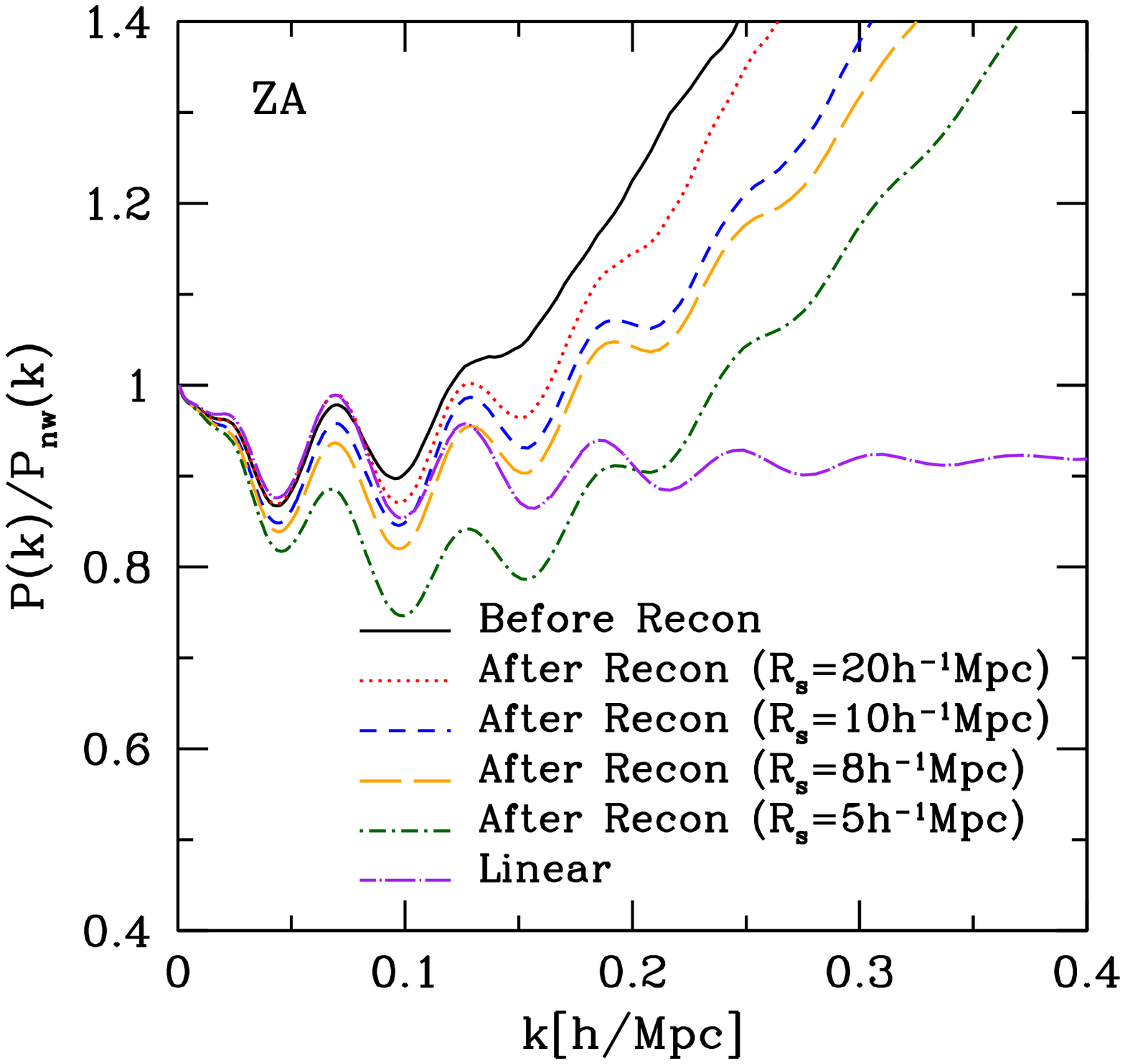}
\caption{Comparison of the one-loop components $P_{22}(k)$ (upper
  left), $P_{13}(k)$ (lower left), their sum (upper right), and the
  matter power spectrum including the one-loop components normalized
  with the no-wiggle components $P_{\rm nw}$ (lower right).  The solid
  line denotes the result before reconstruction while other lines show
  the results after reconstruction with different $R_s=20h^{-1}$Mpc
  (dotted lines), $10h^{-1}$Mpc (short-dashed lines), $8h^{-1}$Mpc
  (long-dashed lines) and $5h^{-1}$Mpc (dot-dashed lines). The
  displacement field in reconstruction is computed with the inverse ZA
  of the smoothed density field with a Gaussian filter at different
  smoothing scales $R_s$ (eq. \ref{eq:sk_ZA}).}
\label{fig:pkcomp}
\end{center}
\end{figure*}
\begin{figure*}
\begin{center}
\includegraphics[width=8cm]{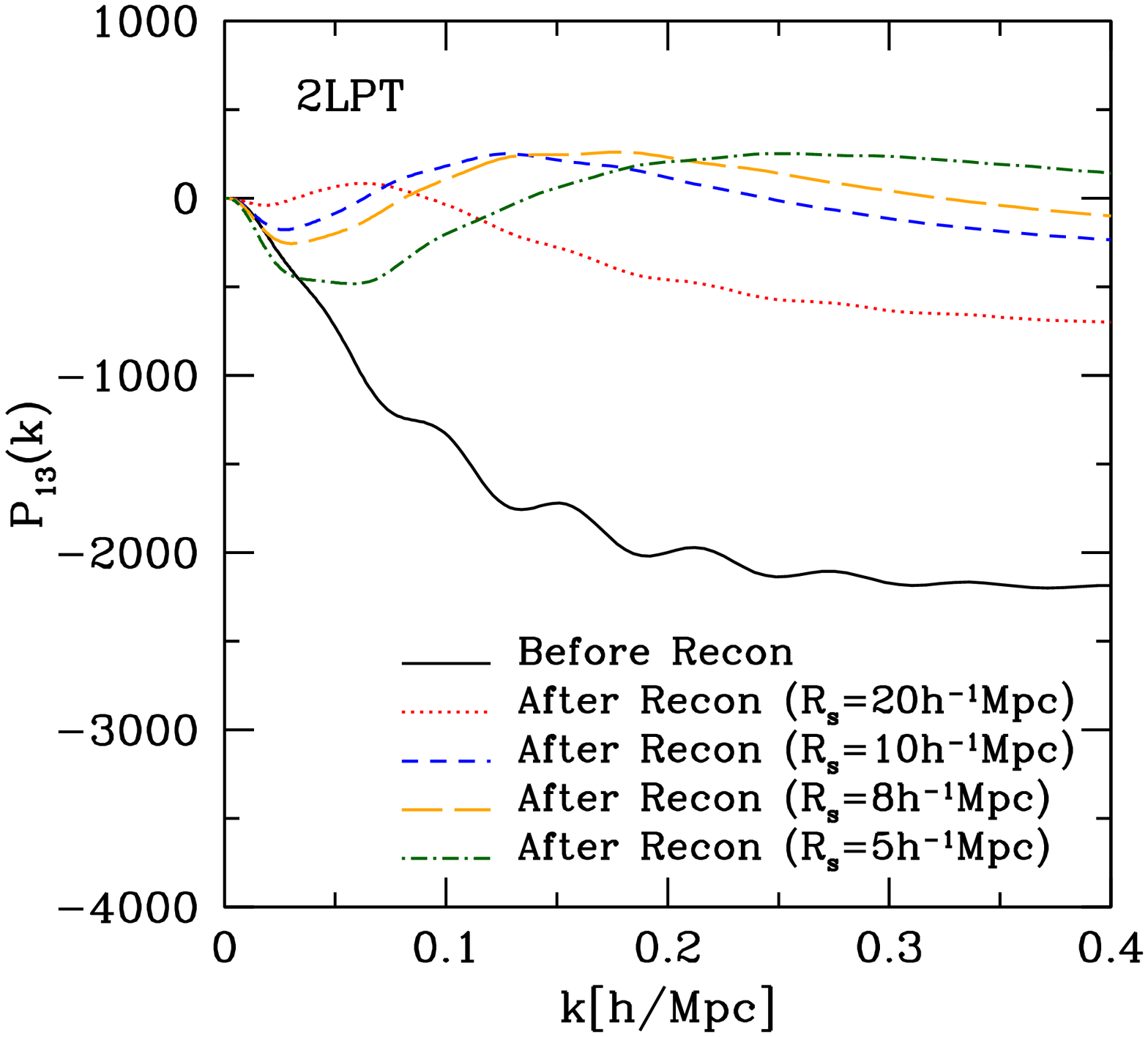}
\includegraphics[width=8cm]{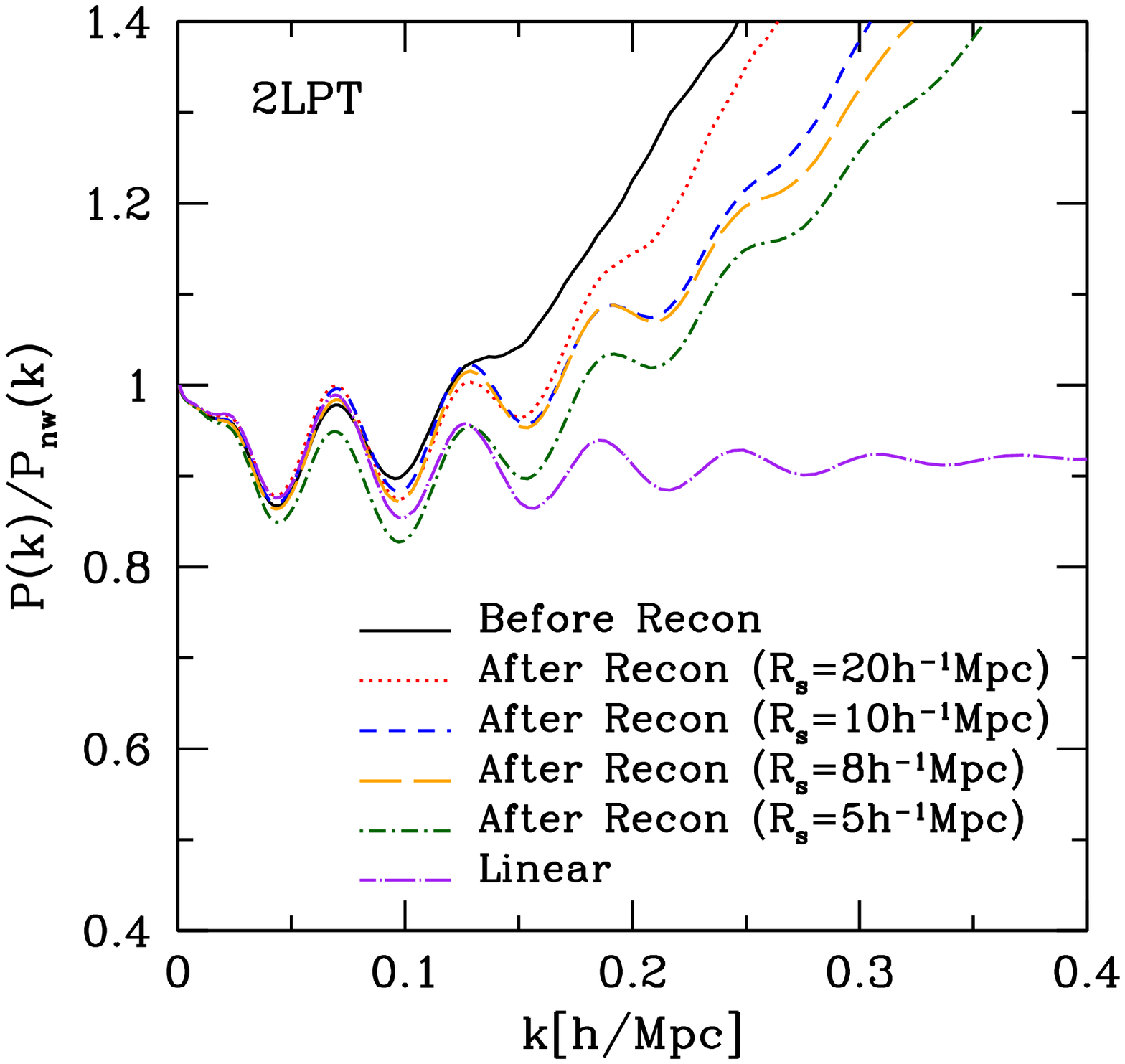}
\caption{Same as the lower two panels in Fig. \ref{fig:pkcomp} but for
  the 2LPT reconstruction.}
\label{fig:pkcomp2}
\end{center}
\end{figure*}
Let's see how the one-loop contributions of the power spectrum change
after reconstruction. Figure \ref{fig:pkcomp} shows the comparison of
the one-loop components $P_{22}(k)$ and $P_{13}(k)$ before and after
reconstruction with different $R_s$. After reconstruction, the
amplitude of the $P_{22}$ term substantially decreases even at high $k$
and the decrement is more significant at smaller $R_s$. This indicates
that the reconstruction significantly reduces the mode-coupling effect due to nonlinear gravity. One can understand how 
$P_{\rm 22}$ decreases from the equation for the $F_2$ kernel
(eq. \ref{eq:F2_recon}).  On very large scales
($W(k)\rightarrow 1$), the final term in the right-hand side cancels
out by reconstruction. This term corresponds to the nonlinearity due to
the shift term $\Psi^{(1)}\cdot \Delta\delta^{(1)}$, which encodes the
motion of the density perturbations due to the gravitational potential
\citep{Sherwin12,Schmittfull15}.

The other one-loop term $P_{13}(k)$, which has in-phase BAO
oscillations but with negative amplitude, contributes to suppress the
intrinsic BAO signature.  The magnitude of the reconstructed
$P_{13}(k)$ also decreases when the smoothed density field is close to
linear, which corresponds to $R_s\simgt 10h^{-1}$Mpc. Since the
amplitude of $P_{13}$ decreases, the BAO signature is substantially
recovered. When the value of $R_s$ is smaller than $10h^{-1}$Mpc,
however, the negative amplitude of $P_{13}(k)$ increases on scales
$k<R_s^{-1}$, which again causes degradation of the BAO signal. The
effect increases as $R_s$ reduces.  This indicates that the behaviour
comes from the nonlinearity in the smoothed density field $\mathbf{s}$
given by the $\mathbf{S}^{(2)}$ term in equation (\ref{eq:F3_recon}).
In the next subsection, we work on this issue by applying the
2nd-order Lagrangian Perturbation Theory (2LPT) reconstruction instead
of ZA.

The upper-right panel in Figure \ref{fig:pkcomp} shows the net
contributions of one-loop terms. Since the one-loop components are
largely canceled out by each other, the net contribution is much
smaller than the amplitude of the individual one-loop terms. When reconstructing
with $R_s\sim 10h^{-1}$Mpc, the net contribution is also smaller than
the unreconstructed case. For $R_s\simlt 10h^{-1}$Mpc, however, the net
contribution becomes negative on $k<R_s^{-1}$ because the negative
amplitude of $P_{13}$ increases. The lower-right panel in Figure
\ref{fig:pkcomp} shows the power spectrum including the one-loop
components normalized by a no-wiggle spectrum $P_{\rm nw}(k)$, calculated from the no-wiggle formula of the
linear spectrum in \citep{EisensteinHu98}. Since the one-loop
components become smaller after reconstruction, one can see that the
reconstructed spectrum approaches the linear one compared with the
unreconstructed spectrum. When $R_s$ is less than 10$h^{-1}$Mpc, the
reconstructed power spectrum is below the linear power spectrum on large
scale ($k<R_s^{-1}$) but again surpasses the linear one as $k$
increases.

\subsection{Reconstruction using 2LPT}
In the previous subsection, we find that the negative amplitude of
$P_{13}$ increases and the deviation from the linear spectrum
increases when the scale $R_s$ of the smoothed density field to be
used for reconstruction is smaller. We confirm that this behaviour comes
from the nonlinearity in the smoothing density field by comparing the
2LPT reconstruction with the ZA reconstruction.

When the shift field is computed based on 2LPT
\citep{Scoccimarro98,Seo10}, it is given in terms of the gravitational potential $\phi$ as
\begin{equation}
\label{eq:shift_2LPT}
\mathbf{s}=-\nabla\phi^{(1)}+\frac{3}{7}\nabla\phi^{(2)}.
\end{equation}
The first term in the right-hand side represents the ZA approximation
and then the potential $\phi^{(1)}$ is computed from the smoothed
density field $\delta^{\rm (smoothed)}$ using the following relation:
\begin{equation}
\nabla^2\phi^{(1)}=\delta^{\rm (smoothed)}.
\end{equation}
The second term in the equation (\ref{eq:shift_2LPT}) corresponds to
the 2LPT correction term and then the potential $\phi^{(2)}$ is
computed from $\phi^{(1)}$ as
\begin{equation}
\label{eq:phi_2LPT}	\nabla^2\phi^{(2)}=\frac12\sum_{i\neq j}
\left\{\phi^{(1)}_{,ii}\phi^{(1)}_{,jj}
	-[\phi^{(1)}_{,ij}]^2\right\},
\end{equation}
where $i,j$ denote $x,y$ and $z$.  Note that the smoothed density
field $\delta^{\rm (smoothed)}$ is not completely linear and thereby
the ZA potential $\phi^{(1)}$ includes the nonlinearity. In order to
remove the 2nd-order nonlinearity, we use a positive sign of the
2nd-order correction.

In Fourier space, the shift field (eq. \ref{eq:sk_ZA}) is then
modified to be
\begin{eqnarray}
\label{eq:sk_2LPT}
\mathbf{\tilde{s}_k}&=&-iW(k)\mathbf{L}^{(1)}\mathbf{(k)}\tilde\delta_\mathbf{k}
+\frac{i}{2}\int\frac{\mathbf{dk}_1 \mathbf{dk}_2}{(2\pi)^3}
\delta_{\rm D}\left(\mathbf{k}_1\mathbf{+k}_2\mathbf{-k}\right) \nonumber \\
&& \times W_1W_2\mathbf{L}^{(2)}(\mathbf{k}_1,\mathbf{k}_2)\tilde\delta_{\mathbf{k}_1}\tilde\delta_{\mathbf{k}_2},
\end{eqnarray}
The second-order kernel for the shift field in the 2LPT reconstruction
is altered to
\begin{equation}
\label{eq:S2_2LPT}
\mathbf{S}^{(2)}(\mathbf{k}_1,\mathbf{k}_2) \rightarrow
\mathbf{S}^{(2)}(\mathbf{k}_1,\mathbf{k}_2)
+W_1W_2\mathbf{L}^{(2)}(\mathbf{k}_1,\mathbf{k}_2).
\end{equation}
On large scales where $W_1$ and $W_2$ are close to unity, the
$\mathbf{L}^{(2)}$ term included in $\mathbf{S}^{(2)}$ is completely canceled out and
thereby the nonlinearity decreases. One of the one-loop terms, $P_{22}$,
remains the same; however, the other term $P_{13}$ changes to
\begin{eqnarray}
P_{13}^{\rm (rec,2LPT)}&(k)&=P_{13}^{\rm (rec)}(k)+
\frac{k^3}{4\pi^2}P_L(k)\int_0^{\infty}drP_L(kr) \nonumber \\
&\times &\int^{1}_{-1}dx
\frac{6r^2(1-rx)(1-x^2)}{7(1+r^2-2rx)}W_{k}W_{kr}.
\end{eqnarray}

Figure \ref{fig:pkcomp2} shows the results of $P_{13}(k)$ and the
1-loop SPT power spectrum by reconstruction with 2LPT. When
reconstructing with 2LPT, the negative amplitude that appeared at
small $k$ becomes smaller in magnitude and as a result the power
spectrum is closer to the linear spectrum than when reconstructing
using ZA. The difference from the reconstruction in ZA comes from the
second-order kernel for the shift field $\mathbf{S}^{(2)}$ in the
equation (\ref{eq:S2_2LPT}). Our result indicates that the
nonlinearity in the smoothed density field $\mathbf{S}^{(2)}$ causes
the negative amplitude of $P_{13}(k)$ on small $k$.

\section{Comparison of Perturbation Theory with N-body simulations}
\label{sec:results}
We run N-body simulations to generate the reconstructed matter density
field. We compare the perturbative formula of matter power spectrum
and bispectrum derived in the previous section with the numerical
results to study the range of scales in which PT works.
\subsection{N-body simulations}
We use the publicly available cosmological N-body simulation code
Gadget-2 \citep{Springel05}. The mass particles are initially distributed
using the 2LPT code \citep{Crocce06b} with Gaussian initial conditions
at the initial redshift of $49$.  The simulation is performed in a
periodic cubic box with side length $1h^{-1}$Gpc, with $800^3$ 
particles each of mass $1.3\times
10^{11}h^{-1}M_\odot$.

The simulated mass density field $\delta$ is evaluated on an $800^3$ cubic lattice,
using Clouds-in-Cell (CIC) assignment. We compute the shift field $\mathbf{s}(\mathbf{x}_i)$ at each
grid point $\mathbf{x}_i$ as follows: Fourier transforming $\delta(\mathbf{x}_i)$ using the
Fast Fourier transform (FFT) method, multiplying $\delta(\mathbf{k})$ by
$-i\mathbf{k}/k^2 W(kR_s)$, and transforming back to real space. Each
mass particle position $\mathbf{x}$ is shifted to $\mathbf{x+s(x)}$
where $\mathbf{s(x)}$ is linearly interpolated from the shift vectors
at the neighboring pixels, and we recompute the shifted density field
$\delta^{\rm (d)}$. The shifted random (uniform grid) field
$\delta^{\rm (s)}$ is obtained by shifting each grid position by
$\mathbf{s}(\mathbf{x}_i)$. Matter power spectra are computed in
Fourier space by linearly binning with $\Delta
k=0.01h$/Mpc. The calculation of bispectra is also performed in
Fourier space focused on some specific triangle configurations:
$(k_1,k_2)=(0.05,0.1), (0.1,0.2), (0.15, 0.3)$ and $(0.2,0.4)$ in unit
of $h^{-1}$Mpc where the binning width of $k_1$ and $k_2$
$0.01h^{-1}$Mpc and their opening angle $\theta$ varies from 0 to
$\pi$ with a binning width of $\pi/15$. We run 10 realizations to
evaluate the errors of the power spectrum and bispectrum. When
comparing PT with the numerical results, the CIC pixel window function is added to the Gaussian smoothing $W(k)$ in the calculation of PT.

\subsection{Power Spectrum}
\begin{figure*}
\begin{center}
\includegraphics[width=16cm]{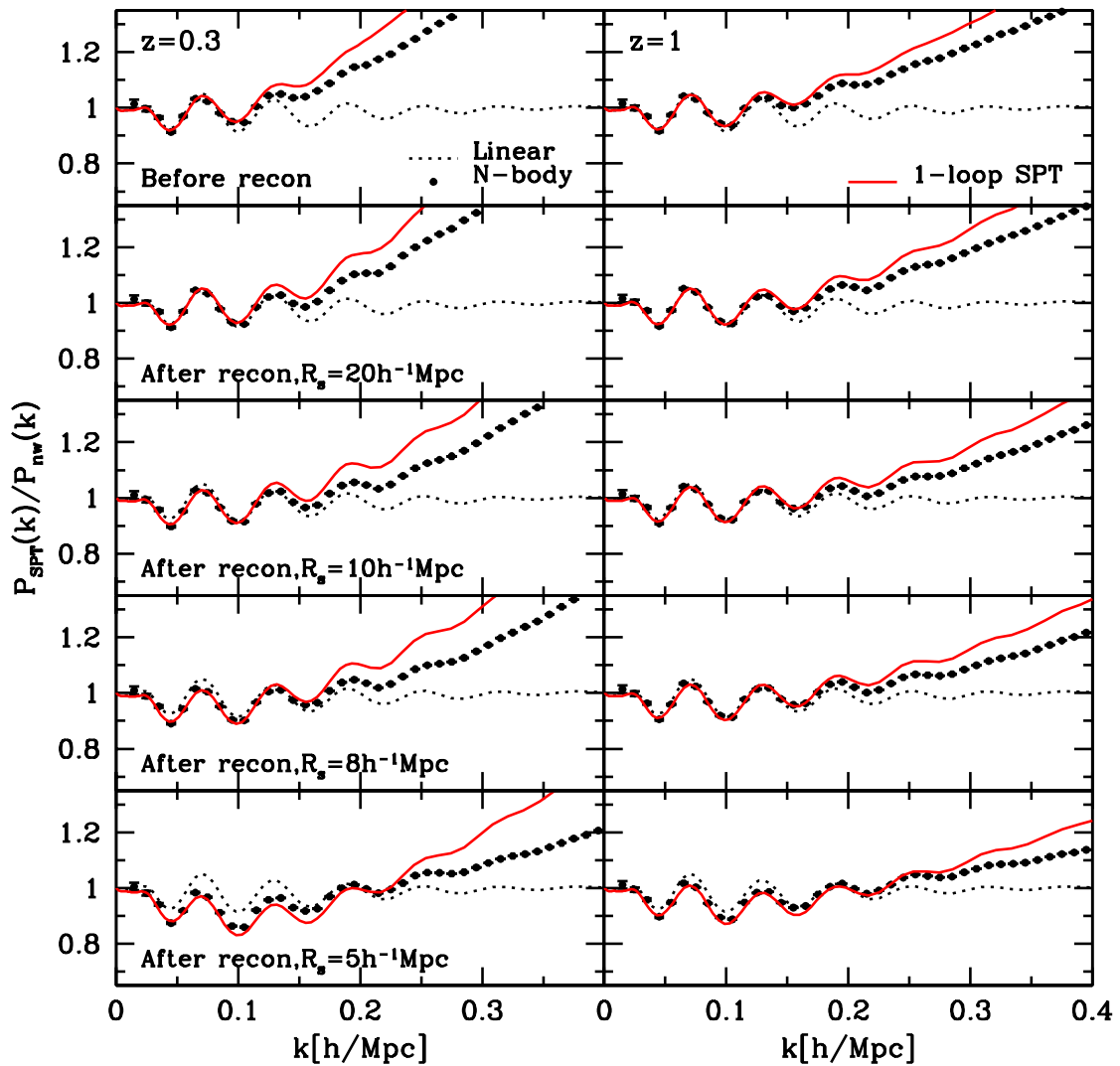}
\caption{Matter power spectrum computed from N-body simulations
  (filled circles) before reconstruction (top) and after
  reconstruction with different $R_s=20, 10, 8$ and $5h^{-1}$Mpc (from
  second-top to bottom). The output redshift is $z=0.3$ (left) and
  $z=1$ (right). The matter power spectra are normalized with the
  no-wiggle spectra at the corresponding redshift. The error-bars
  represent the $1\sigma$ dispersion of our simulation results. Lines
  represent the linear power spectrum (dotted lines) and 1-loop SPT
  (solid lines).}
\label{fig:pk_stdPT}
\end{center}
\end{figure*}
\begin{figure*}
\begin{center}
\includegraphics[width=16cm]{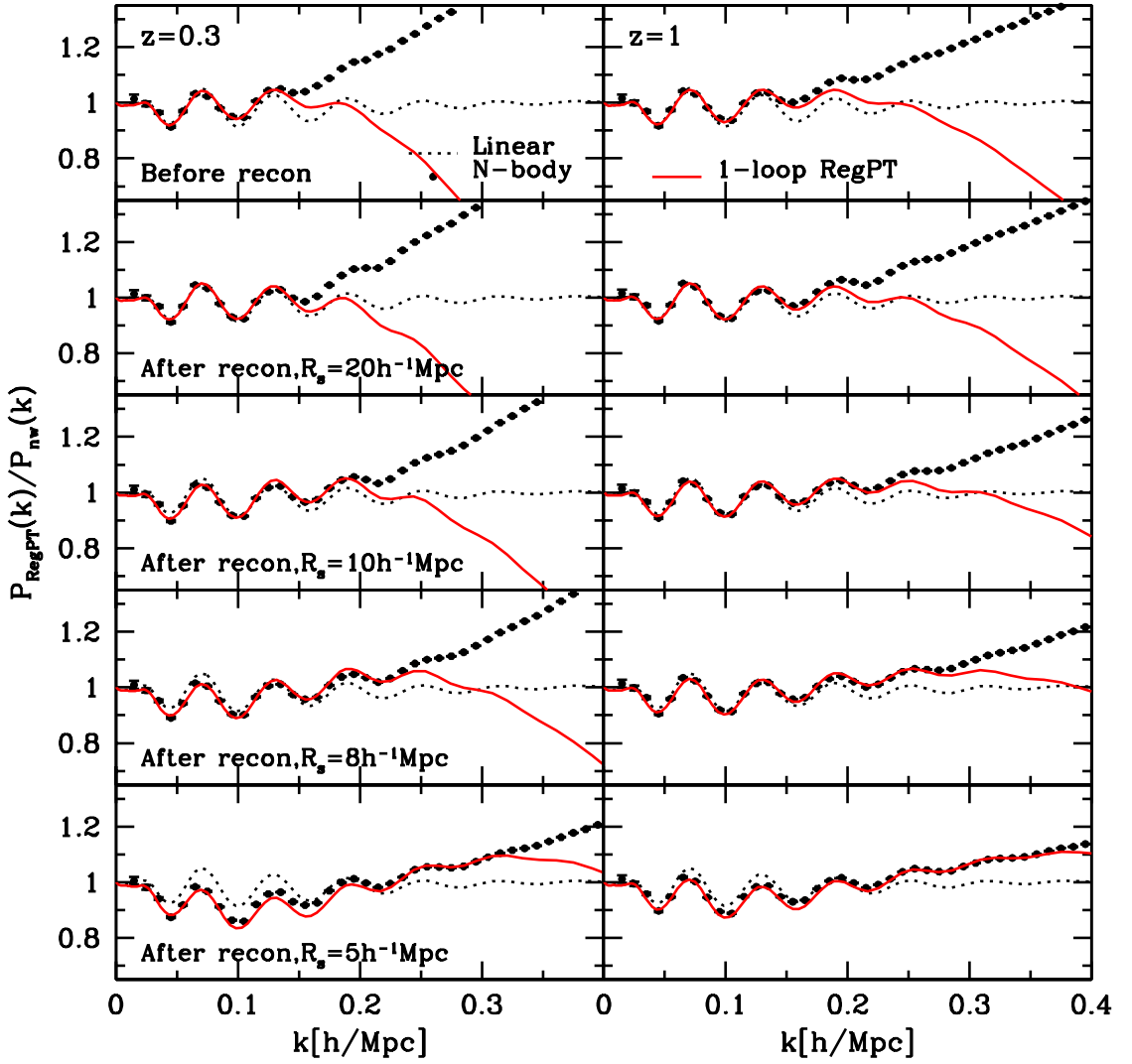}
\caption{Same as Fig. \ref{fig:pk_stdPT} but for the comparison of 
  the one-loop RegPT predictions.}
\label{fig:pk_RegPT}
\end{center}
\end{figure*}
\begin{figure*}
\begin{center}
\includegraphics[width=16cm]{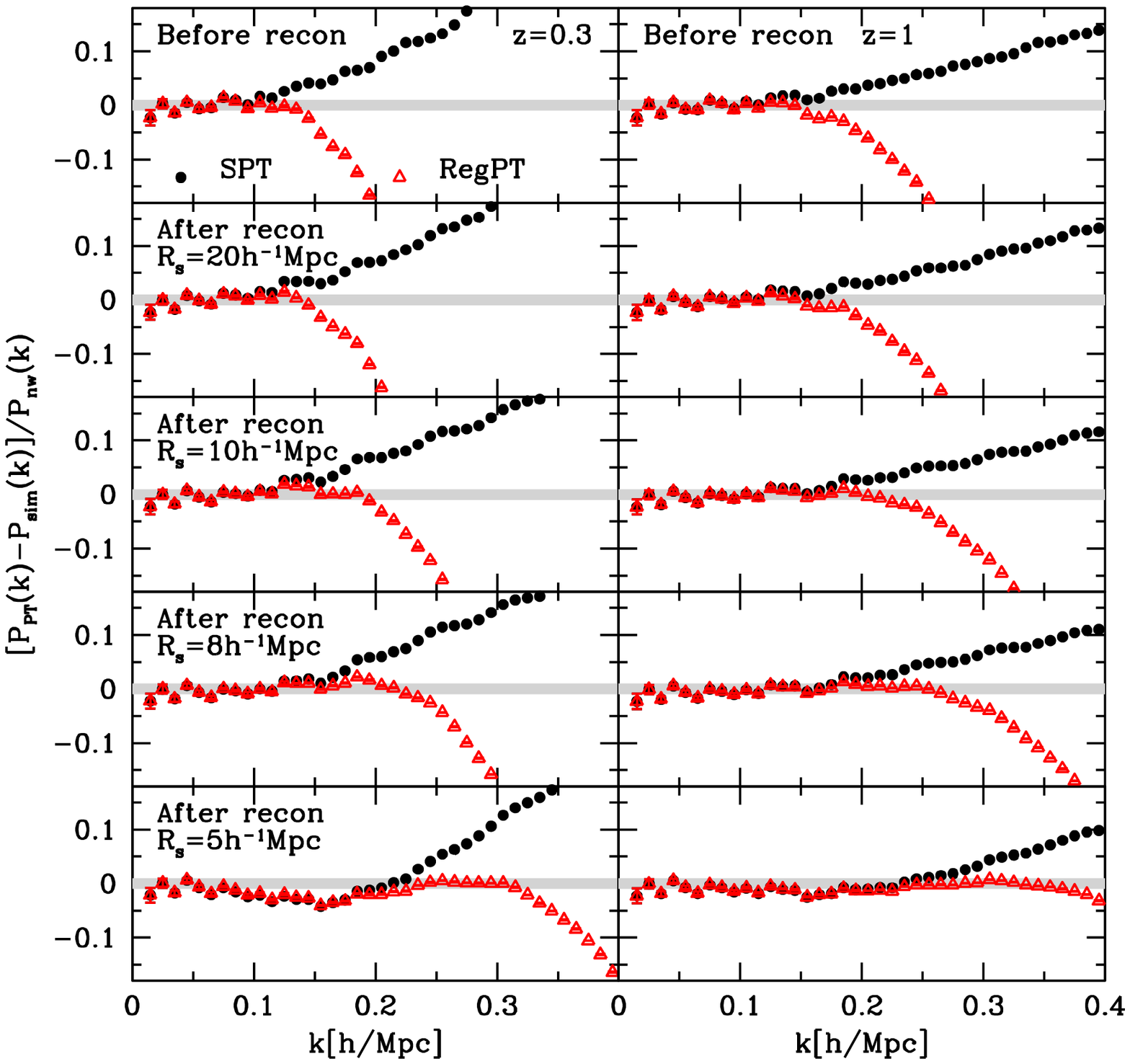}
\caption{Deviation of the one-loop PT of matter power spectrum (SPT
  with filled circles and RegPT with open triangles) from N-body
  results before reconstruction (top) and after reconstruction with
  different $R_s$ from $20, 10, 8$ and $5h^{-1}$Mpc (second top to
  bottom). The redshift is $z=0.3$ (left) and $z=1$ (right). The
  shaded area indicates $\pm$1\% range of the deviation.}
\label{fig:dpk}
\end{center}
\end{figure*}

Figure \ref{fig:pk_stdPT} shows the comparison of the one-loop SPT
with the N-body results averaged over 10 realizations at $z=0.3$
(left) and $z=1$ (right). The top panels are the results before
reconstruction and the other panels show the results after
reconstruction with different $R_s$. Before reconstruction, the
one-loop SPT is in good agreement with the N-body results in the
weakly nonlinear regime ($k\simlt 0.1h^{-1}$Mpc) with errors around
one percent, while the SPT overestimates the matter power at higher
$k$. This behaviour is consistent with the previous works
\citep[e.g.,][]{JeongKomatsu06,Matsubara08a}. After reconstruction,
the simulated matter power spectrum approaches the linear spectrum for
$R_s\simgt 10h^{-1}$Mpc. We find that one-loop SPT agrees with the
N-body results after reconstruction in the weakly nonlinear regime,
while PT still overestimates in nonlinear regime. The agreement of the
one-loop PT with the N-body results is better up to larger $k$ than
the prereconstruction case when $R_s$ is chosen to be
$8-10h^{-1}$Mpc. This indicates that the nonlinear gravity effect
becomes smaller by reconstruction and thereby the range of scales on
which PT holds extends to larger $k$.  As $R_s$ is larger, the
reconstruction becomes ineffective because the reconstuction only
corrects the bulk motion at scales larger than around $R_s$. The limit
that $R_s$ goes infinity corresponds to no reconstruction.

When $R_s$ is smaller than $10h^{-1}$Mpc, the deviation from the
linear spectrum increases at small $k$. This behaviour is consistent
with the predictions from PT that the net ampltitude of the one-loop
terms increase on large scale. When $R_s=5h^{-1}$Mpc, there is a
deviation between PT and the numerical results on relatively large
scale ($k\sim 0.1h^{-1}$Mpc). This comes from that the nonlinearity in
the smoothed density field becomes significant for small $R_s$ and
thereby PT breaks down even for large scale modes, as discussed in the
previous section. Nevertheless, the deviation becomes smaller at
higher redshift and then the agreement between PT and the numerical
results holds up to much larger $k$ than before reconstruction.

We also apply the regularized power spectrum (RegPT) to describe the
power spectrum for the reconstructed field. RegPT is an improved PT in which
the SPT is reorganized with the multi-point propagators
\citep{Crocce06a,Bernardeau08,Taruya12}. Analytical expression up to
one-loop order is given by \citep{Taruya12} as
\begin{eqnarray}
P(k;\eta)&=&[\Gamma^{(1)}_{\rm reg}(k;\eta)]^2P_0(k) \nonumber \\
&+&2\int \frac{\mathbf{dq}}{(2\pi)^3}[\Gamma^{(2)}_{\rm reg}(\mathbf{q},\mathbf{k-q};\eta)]^2 P_0(q)P_0(|\mathbf{k-q}|), \nonumber \\
\end{eqnarray}
where $\eta\equiv\ln D(t)$ and the multi-point propagators $\Gamma^{(n)}_{\rm reg}$ are
\begin{eqnarray}
\Gamma^{(1)}_{\rm reg}(k;\eta)&=&e^\eta\left[
1+\frac{k^2\sigma_d^2e^{2\eta}}{2}+e^{2\eta}\bar\Gamma^{(1)}_{\rm 1-loop}(k)\right] \nonumber \\
&& \times \exp\left\{-\frac{k^2\sigma_d^2e^{2\eta}}{2}\right\}, \\
\Gamma^{(2)}_{\rm reg}(\mathbf{q},\mathbf{k-q};\eta)&=&e^{2\eta}F_2(\mathbf{q},\mathbf{k-q})\exp\left\{-\frac{k^2\sigma_d^2e^{2\eta}}{2}\right\}, \nonumber \\
\end{eqnarray}
and 
\begin{equation}
\bar\Gamma^{(1)}_{\rm 1-loop}(k)=3\int\frac{\mathbf{dq}}{(2\pi)^3}F_3(\mathbf{q},\mathbf{-q},\mathbf{k})P_0(q).
\end{equation}
We obtain the RegPT formula for the reconstructed spectrum by
replacing the Eulerian kernels $F_n$ with $F_n^{\rm (rec)}$
(eq. \ref{eq:F2_recon} for $n=2$ and eq. \ref{eq:F3_recon} for
$n=3$). The $\sigma_d$ represents the dispersion of the displacement
field
\begin{equation}
\label{eq:sigmad}
\sigma_d^2(k)=\int_0^{k_\Lambda(k)}\frac{\mathbf{dq}}{6\pi^2}P_L(\mathbf{q}),
\end{equation}
where the running UV cutoff $k_\Lambda(k)$ is set to be $k/2$ by
\cite{Taruya12}. Since the Lagrangian displacement becomes effectively
smaller by reconstruction \citep{Padmanabhan09}, the value of
$\sigma_d^2(k)$ should be smaller after reconstruction. We then set
$(\sigma_d^2(k))^{\rm recon}=b\sigma_d^2(k)$ with $b$ treated as a
free parameter to fit the simulated spectrum.

Figure \ref{fig:pk_RegPT} shows the comparison of the one-loop RegPT
power spectrum with the simulated results at $z=0.3$ (left) and $z=1$
(right).  Similar to Figure \ref{fig:pk_stdPT}, the top panels show the
results before reconstruction and the other panels are the results
after reconstruction with different $R_s$. RegPT describes the matter
power spectrum better than the SPT at one-loop level.  On nonlinear
scales, RegPT underestimates the matter power spectrum due to the
exponential damping for the dispersion of the displacement
(eq. \ref{eq:sigmad}). After reconstruction, we find that the smaller
value of $b$ is better fitted to the N-body results at smaller $R_s$:
$b=1$ for $R_s=20h^{-1}$Mpc, $b=0.7$ for $R_s=10h^{-1}$Mpc, $b=0.5$
for $R_s=8h^{-1}$Mpc, and $b=0.3$ for $R_s=5h^{-1}$Mpc. When $R_s$ is
around $8h^{-1}$Mpc, the agreement between one-loop RegPT and the
N-body results reaches up to the third BAO peak ($k\sim 0.18h$/Mpc)
at $z=0.3$ and the fourth BAO peak ($k\sim 0.24h$/Mpc) at $z=1$.

Figure \ref{fig:dpk} shows the deviation of the power spectrum between
PT and N-body simulations normalized with the no-wiggle components
[$(P_{\rm PT}(k)-P_{\rm sim}(k))/P_{\rm nw}(k)$]. The gray shaded area
represents the 1 percent range of the deviation.  Before
reconstruction, the agreement of the one-loop RegPT with the
simulations is good up to around $0.1h^{-1}$Mpc within 1\% error. After
reconstruction, their agreement extends up to  $0.2-0.25\,h$Mpc$^{-1}$ when
$R_s$ chosen to be $8-10h^{-1}$Mpc. Our results indicate that the
higher-order mode-coupling beyond the one-loop order becomes smaller
by reconstruction and then the agreement of PT better holds to
higher $k$.

\begin{figure*}
\begin{center}
\includegraphics[width=16cm]{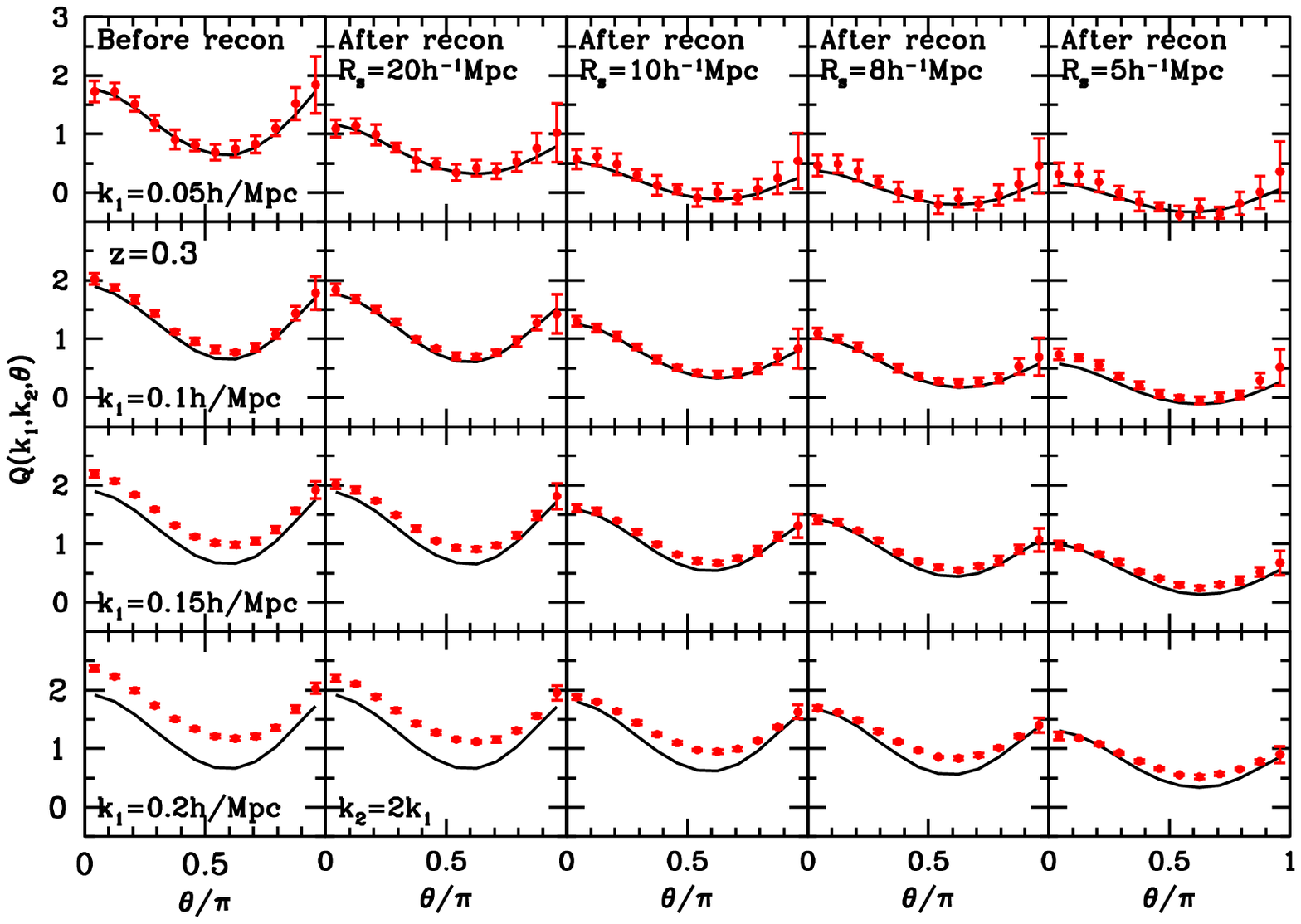}
\includegraphics[width=16cm]{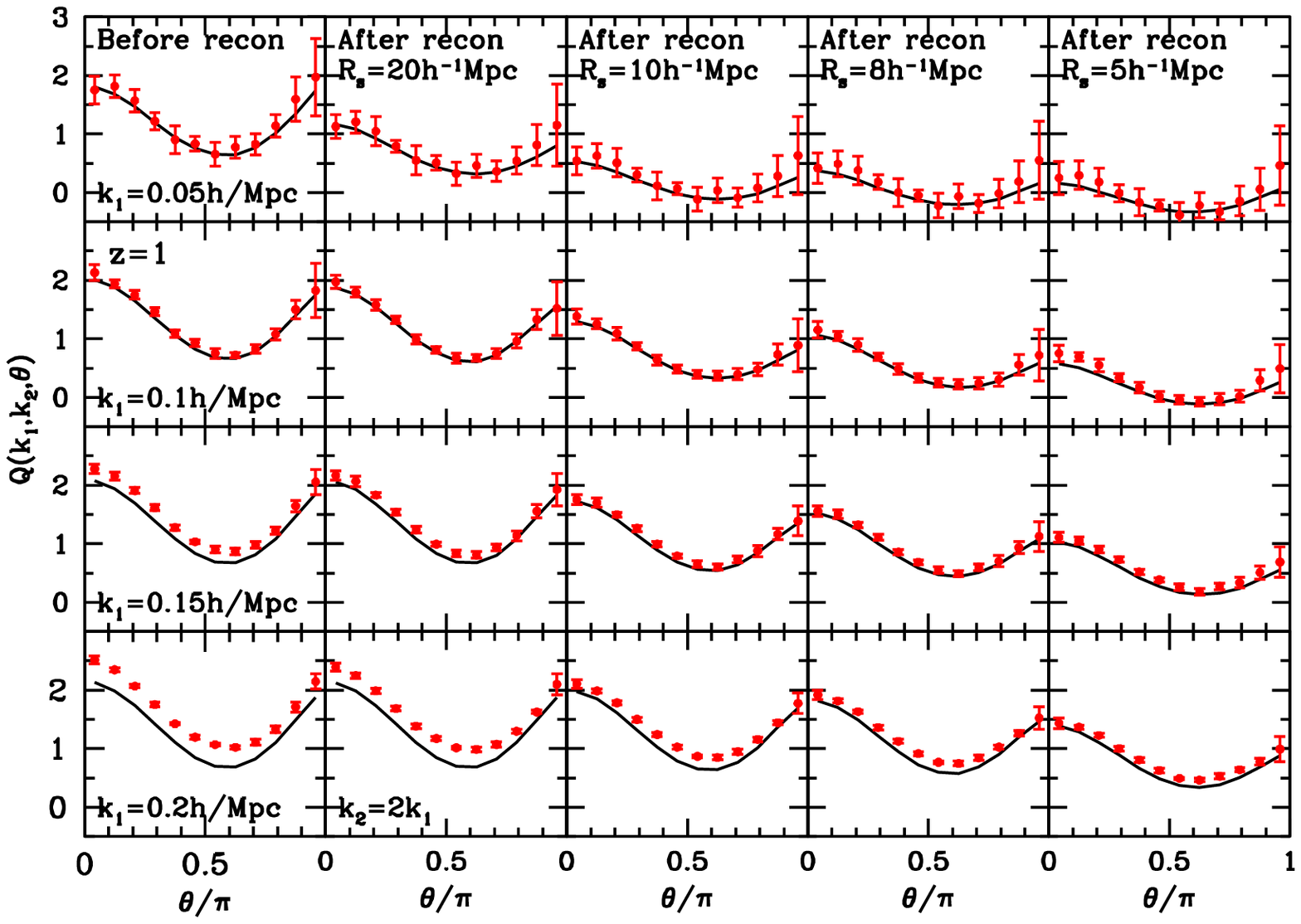}
\caption{Reduced bispectrum $Q$ (eq. \ref{eq:Q}) for real-space dark
  matter field before and after ZA reconstruction with different
  smoothing scale $R_s=20, 10, 8$ and $5h^{-1}$Mpc from left to
  right. The redshift is 0.3 (Upper) and 1 (Lower). The configuration
  of the triangle is fixed to be $k_2/k_1=2$ with $k_1=0.05, 0.1,
  0.15, 0.2h$/Mpc (from top to bottom) and the opening angle
  $\theta\equiv\arccos(\mathbf{k_1\cdot k_2}/k_1k_2$) is varied. Lines
  represent the tree-level PT (eqs. [\ref{eq:bis_unrecon}] and
  [\ref{eq:bis_recon}]) and the symbols are N-body results. The
  error-bars represent the $1\sigma$ dispersion of the N-body results
  with 1$(h^{-1}{\rm Gpc})^3$ volume.}
\label{fig:bis_recon}
\end{center}
\end{figure*}

\subsection{Bispectrum}
\label{subsec:bispec}
We also compare the perturbative formula of the matter bispectrum at
tree-level order (eqs. [\ref{eq:bis_unrecon}] and
[\ref{eq:bis_recon}]) with the N-body results. Figure
\ref{fig:bis_recon} shows the comparison of the reduced bispectrum $Q$
between PT and N-body simulations at $z$=0.3 and 1:
\begin{equation}
\label{eq:Q}
Q(k_1,k_2,k_3)\equiv 
\frac{B(k_1,k_2,k_3)}{P(k_1)P(k_2)+P(k_2)P(k_3)+P(k_3)P(k_1)}.
\end{equation}
We focus on the triangle configuration of $k_2/k_1=2$ varying the
opening angle $\theta\equiv \arccos(\mathbf{k_1\cdot k_2}/k_1/k_2)$
with different values of $k_1$ [$h$/Mpc] =$\{0.05, 0.1, 0.15, 0.2
\}$. The overall amplitude of the reconstructed bispectrum decreases
by reconstruction as $R_s$ decreases. This is because the
mode-coupling effect appeared in $F_2$ kernel is significantly reduced
by reconstruction as shown in Figure \ref{fig:pkcomp}. The
reconstructed matter density field is thereby closer to be
Gaussian. We find that the agreement between the tree-level PT and the
N-body results is better after reconstruction as $R_s$ is smaller upto
$R_s=8-10h^{-1}$Mpc. At smaller $R_s$, however, the nonlinearity in
the smoothed density fields again cause the deviation of the
bispectrum on small $k$ as well as that of the power spectrum.  The
deviation should be improved by taking into account the one-loop
components in the bispectrum \citep[e.g.,][]{Sefusatti09}, however, we
leave this work for the future.

\subsection{Reconstruction using 2LPT}
\begin{figure*}
\begin{center}
\includegraphics[width=16cm]{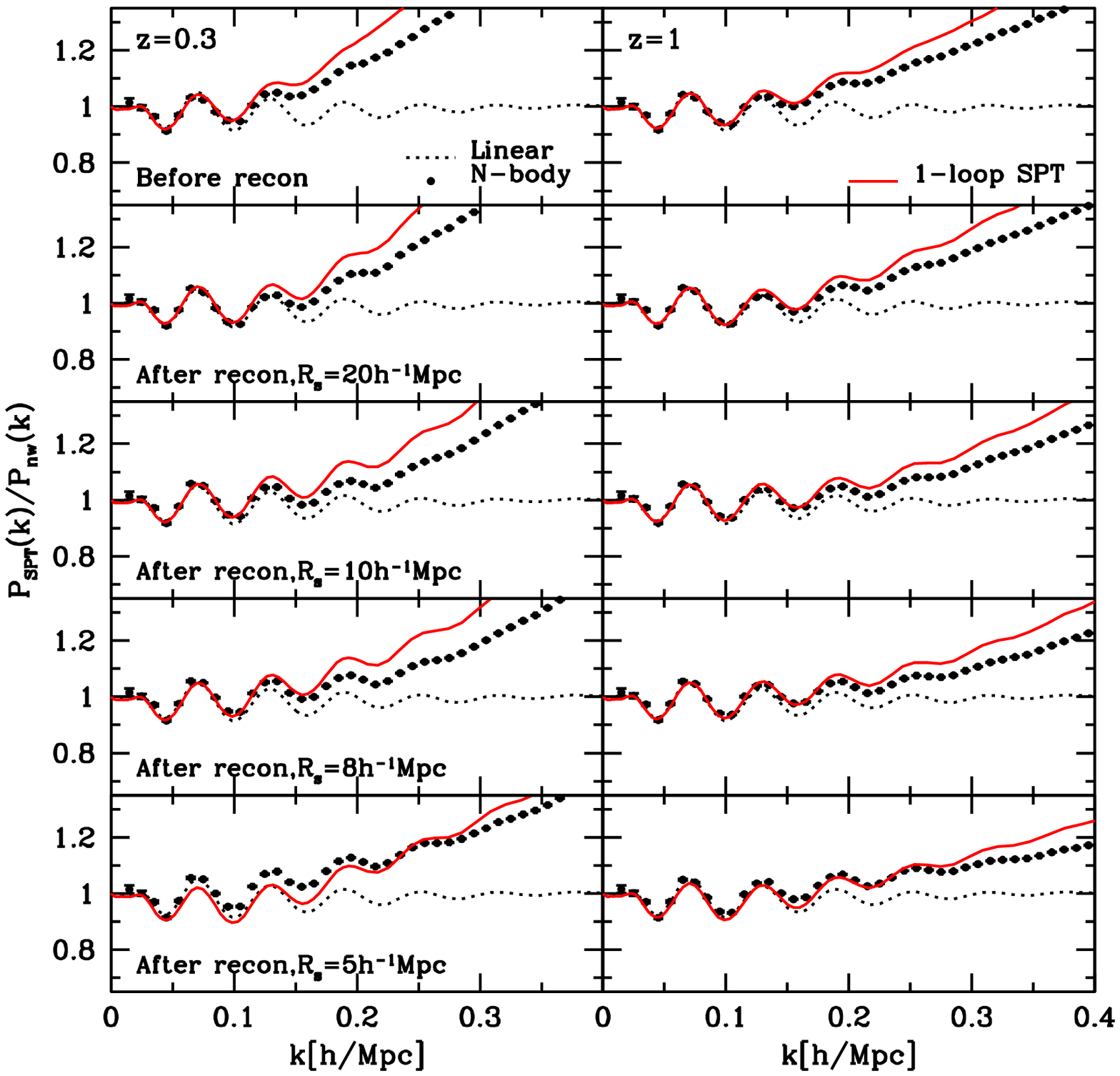}
\caption{Same as Fig. \ref{fig:pk_stdPT} but for the reconstructed
  power spectrum using 2LPT instead of ZA.}
\label{fig:pk_2LPT}
\end{center}
\end{figure*}
In this subsection, we show the reconstructed power
spectrum using 2LPT instead of ZA. As shown in Figure
\ref{fig:pkcomp2}, the negative amplitude due to the nonlinearity in
the smoothing density field is alleviated by including the 2LPT
correction term and the reconstructed spectrum is closer to the
linear spectrum on small $k$.  In order to confirm this, we use the
N-body simulations to compute the reconstructed spectrum using
2LPT. We numerically compute the 2LPT correction term
(eq. \ref{eq:phi_2LPT}) by following the prescription written in
Appendix D2 of \cite{Scoccimarro98}. In Fourier space, we compute
the $\phi^{(1)}_{,ij}$ ($i,j$=1,2,3) terms by multiplying
$\tilde\delta(k)$ by $-(k_ik_j/k^2) W(kR_s)$, transforming back to
real space and computing the 2LPT source term in the right-hand side
of eq. \ref{eq:phi_2LPT}.

Figure \ref{fig:pk_2LPT} shows the comparison of PT with the
N-body results when the matter density field is reconstructed using
2LPT. We confirm that the N-body results approach the linear
spectrum for small $k$. This indicates that the nonlinearity in the
smoothed density field can be partially canceled by using 2LPT. 
However, we find that the agreement with PT is not improved significantly 
by using 2LPT. 

\section{Summary and Conclusions}
\label{sec:summary}
We derive the one-loop order perturbative formula of the real-space
matter power spectrum applied with the standard Lagrangian BAO
reconstruction technique, in which the objects are displaced by the
inverse ZA of the density field smoothed at $R_s$. We find that both
of the next-leading one-loop terms $P_{22}$ and $P_{13}$ decrease in
magnitude by reconstruction and thereby the reconstructed spectrum
approaches the linear spectrum as long as the smoothed density
field is close to linear, i.e, the smoothing scale $R_s\simgt
10h^{-1}$Mpc. Compared with N-body simulations, we find that our PT
formula works also after reconstruction when the field is weakly
non-Gaussian. When the smoothing scale $R_s$ for the displacement
field is smaller than $\sim 10h^{-1}$Mpc, however, we find 
deviations from the linear power spectrum on large scales (small $k$).
By using the 2LPT approximation in reconstruction instead of
inverse ZA, we find that this comes from the nonlinearity in the smoothed density
field.  Compared with the numerical simulations, we confirm that
the numerical results show the behavior consistent with PT
predictions in a wide range of $R_s$. We find that the agreement
between PT and numerical simulations is better after
reconstruction with $R_s\sim 8-10h^{-1}$Mpc. 
We also apply the RegPT to describe the nonlinearity
in the reconstructed density field with the damping factor treated as
a free parameter. We find that after reconstruction RegPT describes
the matter power spectrum up to the third BAO peak ($k\sim
0.18h$/Mpc) at $z=0.3$ and the fourth BAO peak ($k\sim
0.24h$/Mpc) at $z=1$ even at one-loop order.

We also investigate the bispectrum for the reconstructed field in
a perturbative manner. The amplitude of the bispectrum is found to
decrease significantly after reconstruction. This is consistent with
the fact that the nonlinear mode-coupling effect weakens by
reconstruction and thereby the non-Gaussianity becomes smaller after
reconstruction. We find that the tree-level PT agrees with the
simulation results in the weakly nonlinear regime, and the deviation between
PT predictions and simulation results becomes smaller after
reconstruction.

We confirm that the mode-coupling effect due to the nonlinear gravity
becomes smaller by reconstruction in a perturbative approach. This may
be practically useful to extract cosmological information from higher
$k$ modes than the previous work without reconstruction
\citep[e.g.,][]{Reid12,Beutler13}. Two important caveats should be
borne in mind: in order to apply our perturbative formula for the
actual galaxy surveys, we have to take into account the other
nonlinear effects such as redshift-space distortion and also galaxy
bias. These nonlinear effects can also be incorporated in a
perturbative manner
\cite[e.g.,][]{Heavens98,Scoccimarro04,Matsubara08b,Taruya10}. Although
there is evidence that galaxies are unbiased on large scales at low
redshift \cite{Verde02}, this is not true at high redshift
(e.g. \cite{Chiang15}), and it depends on the sample. We will present
the application of our work to redshift-space clustering and also to
the halo density fields in the next paper.

It is also interesting to study how the information of primordial
density field is extracted from the reconstructed density field.
Since the reconstructed field effectively approaches the initial
density field, the reconstructed field is better suited to constrain
the primordial information such as primordial non-Gaussianity
\citep{Scoccimarro04,Sefusatti09}. This is true unless the signature
of the primordial information is not canceled by reconstruction.

\begin{acknowledgments}
We thank Florian Beutler for useful comments. CH is supported by
MEXT/JSPS KAKENHI Grant Numbers 16K17684. KK is supported by the UK
Science and Technologies Facilities Council grants ST/N000668/1 and
the European Research Council through grant 646702 (CosTesGrav).
\end{acknowledgments}

\appendix

\section{Derivation of the Eulerian kernel for the reconstructed
  field}
\label{sec:app}
In this Appendix, we derive a perturbative formula of the
reconstructed matter density field in real space.  We expand the
equation [\ref{eq:reconfield}]) up to third order to derive the
second- and third-order Eulerian kernels $F_n^{\rm (rec)} (n=2,3)$. We
take into account the higher-order terms of the shift field
$\mathbf{s(x)}$ and the difference of Eulerian and Lagrangian
positions in the shift field, $\mathbf{s(x)-s(q)}$.

\subsection{$F_2^{\rm (rec)}$ in ZA}
As shown in equation (\ref{eq:sx_pb}), the leading order of the
difference of the shift field evaluated at Eulerian and Lagrangian
positions $\Delta
\mathbf{s}=\mathbf{s}(\mathbf{x})-\mathbf{s}(\mathbf{q})$ starts at
second order:
\begin{eqnarray}
\Delta\mathbf{s}^{(2)}&=&(\mathbf{\Psi}^{(1)}\mathbf{\cdot\nabla})\mathbf{s}^{(1)}, \\
\Delta\mathbf{s}^{(3)}&=&(\mathbf{\Psi}^{(2)}\mathbf{\cdot\nabla})\mathbf{s}^{(1)}
+(\mathbf{\Psi}^{(1)}\mathbf{\cdot\nabla})\mathbf{s}^{(2)} \nonumber \\
&&+\frac12(\Psi^{(1)}_i\Psi^{(1)}_j\nabla_i\nabla_j)\mathbf{s}^{(1)}.
\label{eq:s3}
\end{eqnarray}
The reconstructed density field given by equation (\ref{eq:reconfield})
is rewritten as
\begin{eqnarray}
\tilde\delta_{\mathbf k}^{\rm (rec)}
&=&\int \mathbf{dq}e^{-i\mathbf{k\cdot q}}
e^{-i\mathbf{k\cdot s(q)}}(e^{-i\mathbf{k\cdot[\Psi(q)+\Delta s(q)]}}-1). \nonumber \\
\end{eqnarray}
The second order of $\tilde\delta^{\rm (rec)}_\mathbf{k}$ is given by
\begin{eqnarray}
\tilde\delta^{\rm (rec)(2)}_\mathbf{k}&=&\int\mathbf{dq}e^{-i\mathbf{k\cdot q}}
\left[(-i\mathbf{k\cdot\Psi}^{(2)}) 
+\frac{1}{2}(-i\mathbf{k\cdot\Psi}^{(1)})^2\right. \nonumber \\
&&\left.+(-i\mathbf{k\cdot s}^{(1)})(-i\mathbf{k\cdot\Psi}^{(1)})
+(-i\mathbf{k}\cdot\Delta\mathbf{s}^{(2)})
\right]. \nonumber \\
\end{eqnarray}
The first two terms are included in the unreconstructed density field
and their symmetrized kernels are respectively given as
\begin{eqnarray}
F_2^{\Psi^{(2)}}&=&\frac12 \mathbf{k\cdot L}^{(2)}(\mathbf{k}_1,\mathbf{k}_2), \\
F_2^{\Psi^{(1)},\Psi^{(1)}}&=&\frac12 (\mathbf{k\cdot L}^{(1)}(\mathbf{k}_1))
(\mathbf{k\cdot L}^{(1)}(\mathbf{k}_2)). 
\end{eqnarray}
The last two terms come from the product of $\mathbf{\Psi}$ and
$\mathbf{s}$, and the leading order of $\Delta\mathbf{s}$ as
\begin{eqnarray}
F_2^{\mathbf{\Psi}^{(1)},\mathbf{s}^{(1)}}&=&
\frac12 [(\mathbf{k\cdot S}^{(1)}(\mathbf{k}_1))(\mathbf{k\cdot L}^{(1)}(\mathbf{k}_2))+{\rm (1~perm.)}], \nonumber \\ \\
F_2^{\Delta\mathbf{s}^{(2)}}
&=&-\frac12 [(\mathbf{k\cdot S}^{(1)}(\mathbf{k}_1))(\mathbf{k}_1\mathbf{\cdot L}^{(1)}(\mathbf{k}_2)) 
+{\rm (1~perm.)}], \nonumber \\ 
\end{eqnarray}
where $\mathbf{S}^{\rm (n)}$ is defined as 
the $n$-th kernel of the shift field $\mathbf{s}$ (eq. \ref{eq:shiftkernel}).
The sum of the above two terms simplifies to
\begin{equation}
F_2^{\mathbf{\Psi}^{(1)},\mathbf{s}^{(1)}}+F_2^{\mathbf{\Delta s}^{(2)}}
= \frac12 [(\mathbf{k\cdot S}^{(1)}(\mathbf{k}_1))+{\rm (1~perm.)}].
\end{equation}
The $F_2^{\rm (rec)}$ then is given by summing over all of the above
components as
\begin{eqnarray}
F_2^{\rm (rec)}(&\mathbf{k}_1&,\mathbf{k}_2)=\frac12
\left[\mathbf{k\cdot L}^{(2)}(\mathbf{k}_1,\mathbf{k}_2)\right. \nonumber \\
&&+(\mathbf{k\cdot L}^{(1)}(\mathbf{k}_1)(\mathbf{k\cdot L}^{(1)}(\mathbf{k}_2)) \nonumber \\
&&+\left.(\mathbf{k\cdot S}^{(1)}(\mathbf{k}_1))+(\mathbf{k\cdot S}^{(1)}(\mathbf{k}_2))\right] \\
&=&
\frac57-\frac{W_1+W_2}{2}+\frac27
\left(\frac{\mathbf{k}_1\mathbf{\cdot k}_2}{k_1k_2}\right)^2 \nonumber \\
&&+\left(\frac{\mathbf{k}_1\mathbf{\cdot k}_2}{2k_1k_2}\right)
\left[\frac{k_2}{k_1}(1-W_1)+\frac{k_1}{k_2}(1-W_2)\right].
\nonumber \\
\end{eqnarray}

\subsection{$F_3^{\rm (rec)}$ in ZA}
The third order of the reconstructed density fluctuation
$\tilde\delta^{\rm (rec)}_\mathbf{k}$ is given by
\begin{eqnarray}
\tilde\delta^{\rm (rec)(3)}_\mathbf{k}&=&\int\mathbf{dq}e^{-i\mathbf{k\cdot q}}
\left[(-i\mathbf{k\cdot\Psi}^{(3)})
+(-i\mathbf{k\cdot\Psi}^{(1)})(-i\mathbf{k\cdot\Psi}^{(2)})\right. \nonumber \\
&&+\frac16 (-i\mathbf{k\cdot\Psi}^{(1)})^3 
+(-i\mathbf{k\cdot\Psi}^{(1)})(-i\mathbf{k\cdot}\Delta \mathbf{s}^{(2)}) 
\nonumber \\
&&+(-i\mathbf{k\cdot}\Delta \mathbf{s}^{(3)}) 
+(-i\mathbf{k\cdot s}^{(1)})
\nonumber \\
&&\times \left\{
(-i\mathbf{k\cdot\Psi}^{(2)}) 
+\frac{1}{2}(-i\mathbf{k\cdot\Psi}^{(1)})^2
+(-i\mathbf{k\cdot}\Delta \mathbf{s}^{(2)})
\right\} \nonumber \\
&&+(-i\mathbf{k\cdot s}^{(2)})(-i\mathbf{k\cdot\Psi}^{(1)})
\nonumber \\
&& \left.
+\frac12 (-i\mathbf{k\cdot s}^{(1)})^2(-i\mathbf{k\cdot\Psi}^{(1)})\right],
\end{eqnarray}
In the following, we derive the third-order kernel corresponding to
each term including the factor associated with the perturbative
expansion. All of the kernels are symmetrized among three wavevectors.

The first three terms are included in the unreconstructed density
field and then each term is written as follows:
\begin{eqnarray}
F_3^{\mathbf{\Psi}^{(3)}}&=&\frac16 \mathbf{k\cdot L}^{(3)}(\mathbf{k}_1,\mathbf{k}_2,\mathbf{k}_3), \\
F_3^{\mathbf{\Psi}^{(1)},\mathbf{\Psi}^{(2)}}&=&
\frac16 [(\mathbf{k\cdot L}^{(1)}(\mathbf{k}_1))(\mathbf{k\cdot L}^{(2)}(\mathbf{k}_2,\mathbf{k}_3)) \nonumber \\
&& + {\rm (2~perms)}], \\
F_3^{\mathbf{\Psi}^{(1)},\mathbf{\Psi}^{(1)},\mathbf{\Psi}^{(1)}}&=&
\frac16 (\mathbf{k\cdot L}^{(1)}(\mathbf{k}_1))(\mathbf{k\cdot L}^{(1)}(\mathbf{k}_2))(\mathbf{k\cdot L}^{(1)}(\mathbf{k}_3)). \nonumber \\
&&
\end{eqnarray}
The fourth term comes from the product of $\mathbf{\Psi}$ and
$\Delta\mathbf{s}$ 
\begin{eqnarray}
\label{eq:1d}
F_3^{\Psi^{(1)},\Delta s^{(2)}}
&=& -\frac16 \left[
(\mathbf{k\cdot S}^{(1)}(\mathbf{k}_1))(\mathbf{k\cdot L}^{(1)}(\mathbf{k}_2))
\right. \nonumber \\
&& \left.\times (\mathbf{k}_1\mathbf{\cdot L}^{(1)}(\mathbf{k}_3)) + {\rm (5~perms)}\right].
\end{eqnarray}
The fifth term comes from $\Delta\mathbf{s}^{(3)}$ which has
the three components given in equation (\ref{eq:s3}) and then the kernel
corresponding to each component is given as follows:
\begin{eqnarray}
\label{eq:1b}
F_3^{\mathbf{\Psi}^{(2)}\mathbf{\cdot\nabla s}^{(1)}}
&=& -\frac16 [
(\mathbf{k\cdot S}^{(1)}(\mathbf{k}_1))(\mathbf{k}_1\mathbf{\cdot L}^{(2)}(\mathbf{k}_2,\mathbf{k}_3))
\nonumber \\ 
&&+ {\rm (2~perms)}], \\
\label{eq:1c}
F_3^{(\mathbf{\Psi}^{(1)}\mathbf{\cdot\nabla})\mathbf{s}^{(2)}}
&=& -\frac16 [(\mathbf{k\cdot S}^{(2)}(\mathbf{k}_1,\mathbf{k}_2))
(\mathbf{k}_{12}\mathbf{\cdot L}^{(1)}(\mathbf{k}_3))
\nonumber \\ 
&& + {\rm (2~perms)}], \\
\label{eq:1f}
F_3^{(\Psi^{(1)}_i\Psi^{(1)}_j\nabla_i\nabla_j)\mathbf{s}^{(1)}}
&=& \frac16 [
(\mathbf{k\cdot S}^{(1)}(\mathbf{k}_1))(\mathbf{k}_1\mathbf{\cdot L}^{(1)}(\mathbf{k}_2)) \nonumber \\
&& \times (\mathbf{k}_1\mathbf{\cdot L}^{(1)}(\mathbf{k}_3))
+ {\rm (2~perms)}].
\end{eqnarray}
The sixth and seventh terms come from the products of the first-order terms of
$\mathbf{s}$ and the second-order terms of $\mathbf{\Psi}$:
\begin{eqnarray}
\label{eq:2a}
F_3^{\mathbf{s}^{(1)},\mathbf{\Psi}^{(2)}}
&=& \frac16 [
(\mathbf{k\cdot S}^{(1)}(\mathbf{k}_1))(\mathbf{k\cdot L}^{(2)}(\mathbf{k}_2,\mathbf{k}_3))
\nonumber \\
&&+ {\rm (2~perms)}], \\
\label{eq:2b}
F_3^{\mathbf{s}^{(1)},\mathbf{\Psi}^{(1)},\mathbf{\Psi}^{(1)}}
&=& \frac16 [
(\mathbf{k\cdot S}^{(1)}(\mathbf{k}_1))(\mathbf{k\cdot L}^{(1)}(\mathbf{k}_2)) \nonumber \\
&&\times (\mathbf{k\cdot L}^{(1)}(\mathbf{k}_3))
+ {\rm (2~perms)}].
\end{eqnarray}
The eighth term is a combination of $\mathbf{s}$ and $\Delta\mathbf{s}$:
\begin{eqnarray}
\label{eq:2c}
F_3^{\mathbf{s}^{(1)},\Delta \mathbf{s}^{(2)}}
&=& -\frac16 [
(\mathbf{k\cdot S}^{(1)}(\mathbf{k}_1))(\mathbf{k\cdot S}^{(1)}(\mathbf{k}_2))
\nonumber \\
&& \times (\mathbf{k}_{12}\mathbf{\cdot L}^{(1)}(\mathbf{k}_3))
+ {\rm (2~perms)}].
\end{eqnarray}
The last two terms are the product of the second order of $\mathbf{s}$:
and $\mathbf{\Psi}^{(1)}$
\begin{eqnarray}
\label{eq:3c}
F_3^{\mathbf{\Psi}^{(1)},\mathbf{s}^{(2)}}
&=& \frac16 [
(\mathbf{k\cdot S}^{(2)}((\mathbf{k}_1,\mathbf{k}_2)) \nonumber \\
&& \times (\mathbf{k\cdot L}^{(1)}(\mathbf{k}_3))
+ {\rm (2~perms)}], \\
\label{eq:3a}
F_3^{\mathbf{\Psi}^{(1)},\mathbf{s}^{(1)},\mathbf{s}^{(1)}}
&=& \frac16 [
(\mathbf{k\cdot S}^{(1)}(\mathbf{k}_1))(\mathbf{k\cdot S}^{(1)}(\mathbf{k}_2)) \nonumber \\
&& \times (\mathbf{k\cdot L}^{(1)}(\mathbf{k}_3))
+ {\rm (2~perms)}].
\end{eqnarray}
Combinations of the above terms lead to some simplifications as follows:
\begin{eqnarray}
\label{eq:sum1}
F_3^{\mathbf{\Psi}^{(2)}\mathbf{\cdot\nabla s}^{(1)}}
&+&F_3^{\mathbf{s}^{(1)},\mathbf{\Psi}^{(2)}}=\frac16
[(\mathbf{k\cdot S}^{(1)}(\mathbf{k}_1)) \nonumber \\
&&\times (\mathbf{k}_{23}\mathbf{\cdot L}^{(2)}(\mathbf{k}_{23})) + {\rm (2~perms)}],  \\
\label{eq:sum2}
F_3^{(\mathbf{\Psi}^{(1)}\mathbf{\cdot\nabla})\mathbf{s}^{(2)}}&+&F_3^{\mathbf{\Psi}^{(1)},\mathbf{s}^{(2)}}
=\frac16 [\mathbf{k\cdot S}^{(2)}(\mathbf{k}_1,\mathbf{k}_2) \nonumber \\
&& +{\rm (2~perms)}], \\
\label{eq:sum3}
F_3^{\mathbf{s}^{(1)},\Delta \mathbf{s}^{(2)}}&+&F_3^{\mathbf{\Psi}^{(1)},\mathbf{s}^{(1)},\mathbf{s}^{(1)}}
=\frac16 [(\mathbf{k\cdot S}^{(1)}(\mathbf{k}_1)) \nonumber \\
&& \times (\mathbf{k\cdot S}^{(1)}(\mathbf{k}_2))+{\rm (2~perms)}], \\
\label{eq:sum4}
F_3^{\mathbf{s}^{(1)},\mathbf{\Psi}^{(1)},\mathbf{\Psi}^{(1)}}&+&
F_3^{\Psi^{(1)},\Delta s^{(2)}}+
F_3^{(\Psi^{(1)}_i\Psi^{(1)}_j\nabla_i\nabla_j)\mathbf{s}^{(1)}} \nonumber \\
&& = \frac16 [
(\mathbf{k\cdot S}^{(1)}(\mathbf{k}_1))(\mathbf{k}_{23}\mathbf{\cdot L}^{(1)}(\mathbf{k}_2)) \nonumber \\
&&\times (\mathbf{k}_{23}\mathbf{\cdot L}^{(1)}(\mathbf{k}_3))
+ {\rm (2~perms)}].
\end{eqnarray}
The third-order kernel for the reconstructed density fields 
after reconstruction are then summarized as
\begin{eqnarray}
F_3^{\rm (rec)}(&\mathbf{k}_1&,\mathbf{k}_2,\mathbf{k}_3)=
F_3(\mathbf{k}_1,\mathbf{k}_2,\mathbf{k}_3) 
\nonumber \\
&&+\frac16 \left[2(\mathbf{k\cdot S}^{(1)}(\mathbf{k}_1))F_2(\mathbf{k}_2,\mathbf{k}_3)\right.
\nonumber \\
&&+(\mathbf{k\cdot S}^{(1)}(\mathbf{k}_1))(\mathbf{k\cdot S}^{(1)}(\mathbf{k}_2))
\nonumber \\
&&\left.+(\mathbf{k\cdot S}^{(2)}(\mathbf{k}_1,\mathbf{k}_2))+{\rm (2~perms.)}
\right],
\label{eq:F3_recon2}
\end{eqnarray}
where $F_3$ is the third-order kernel before reconstruction.  The
first term including the second-order Eulerian kernel comes from the
sum of the equations (\ref{eq:sum1}) and (\ref{eq:sum4}).

When ($\mathbf{k}_1,\mathbf{k}_2,\mathbf{k}_3$)=($\mathbf{k,p,-p}$), each component of 
the third-order kernel becomes
\begin{eqnarray}
\label{eq:F3rec1}
F_3^{\mathbf{\Psi}^{(2)}\mathbf{\cdot\nabla s}^{(1)}}
&+&F_3^{\mathbf{s}^{(1)},\mathbf{\Psi}^{(2)}}=0, \\
\label{eq:F3rec2}
F_3^{(\mathbf{\Psi}^{(1)}\mathbf{\cdot\nabla})\mathbf{s}^{(2)}}
&+&F_3^{\mathbf{\Psi}^{(1)},\mathbf{s}^{(2)}} 
=-\frac{W(k_\ast)}{3}\left(\frac{1-r\mu}{1+r^2-2r\mu}\right) \nonumber \\
&& \times \left(\frac{10r+4r\mu^2-7r^2\mu-7\mu}{7r}\right), \\
\label{eq:F3rec3}
F_3^{\mathbf{s}^{(1)},\Delta \mathbf{s}^{(2)}}&+&F_3^{\mathbf{\Psi}^{(1)},\mathbf{s}^{(1)},\mathbf{s}^{(1)}}
= -\frac{W^2(kr)}{6}\frac{\mu^2}{r^2}, \\
\label{eq:F3rec4}
F_3^{\mathbf{s}^{(1)},\mathbf{\Psi}^{(1)},\mathbf{\Psi}^{(1)}}&+&
F_3^{\Psi^{(1)},\Delta s^{(2)}}+
F_3^{(\Psi^{(1)}_i\Psi^{(1)}_j\nabla_i\nabla_j)\mathbf{s}^{(1)}} \nonumber \\
&& = \frac{W(kr)}{3}\mu^2 \left(1+\frac{1}{r^2}\right), 
\end{eqnarray}
where $r=p/k$, $\mu=\mathbf{k\cdot p}/(kp)$ and $k_\ast\equiv
k(1+r^2-2rx)^{1/2}$.

\subsection{$F_3^{\rm (rec)}$ in 2LPT reconstruction}
When reconstructing using 2LPT approximation instead of ZA, the second-order of the
shift density field is altered as
\begin{eqnarray}
\mathbf{\tilde{s}_k}^{(2)}&=&
-iW(k)\mathbf{L}^{(1)}(\mathbf{k})\tilde\delta_\mathbf{k}^{(2)}
+\frac{iD^2(z)}{2}
\int\frac{\mathbf{dk}_1\mathbf{dk}_2}{(2\pi)^3}
\nonumber \\
&& \times \delta_{\rm D}
\left(\mathbf{k}_1+\mathbf{k}_2-\mathbf{k}\right)
W_1W_2\mathbf{L}^{(2)}(\mathbf{k}_1,\mathbf{k}_2)
\tilde\delta^{\rm (1)}_{\mathbf{k}_1}\tilde\delta^{\rm (1)}_{\mathbf{k}_2},
\nonumber \\
\end{eqnarray}
This corresponds to that the second-order kernel for the shift field
in the 2LPT reconstruction as
\begin{equation}
\mathbf{S}^{(2)}(\mathbf{k}_1,\mathbf{k}_2) \rightarrow
\mathbf{S}^{(2)}(\mathbf{k}_1,\mathbf{k}_2)
+W_1W_2\mathbf{L}^{(2)}(\mathbf{k}_1,\mathbf{k}_2).
\end{equation}
When ($\mathbf{k}_1,\mathbf{k}_2,\mathbf{k}_3$)=($\mathbf{k,p,-p}$),
the equation (\ref{eq:F3rec2}) becomes
\begin{eqnarray}
F_3^{(\mathbf{\Psi}^{(1)}\mathbf{\cdot\nabla})\mathbf{s}^{(2)}}
&+&F_3^{\mathbf{\Psi}^{(1)},\mathbf{s}^{(2)}}
=-\frac{W(k_\ast)}{3} \left(\frac{1-r\mu}{1+r^2-2r\mu}\right) 
\nonumber \\
&& \times \left(\frac{10r+4r\mu^2-7r^2\mu-7\mu}{7r}\right) \nonumber \\
&&+ W(k)W(kr)\frac{(1-r\mu)(1-\mu^2)}{7(1+r^2-2r\mu)}.
\end{eqnarray}

\bibliography{ref}

\end{document}